\documentclass{article}

 \usepackage[final]{nips_2016}

\usepackage[utf8]{inputenc} % allow utf-8 input
\usepackage[T1]{fontenc}    % use 8-bit T1 fonts
\usepackage{hyperref}       % hyperlinks
\usepackage{url}            % simple URL typesetting
\usepackage{booktabs}       % professional-quality tables
\usepackage{amsfonts}       % blackboard math symbols
\usepackage{nicefrac}       % compact symbols for 1/2, etc.
\usepackage{microtype}      % microtypography
\usepackage{amssymb}
\usepackage{graphicx}

\usepackage{amssymb,amsmath,array}
\usepackage{multirow}
\usepackage[colorinlistoftodos]{todonotes}
\usepackage{multirow}
\usepackage{subfig}
\usepackage{upgreek}
\usepackage{color} 

\usepackage{footnote}

\title{A practical guide and software for analysing\\$\;$ pairwise comparison experiments}

\author{
  M. P\'erez-Ortiz\\
  Computer Laboratory\\
 University of Cambridge\\
  Cambridge, United Kingdom \\
  \texttt{mp867@cam.ac.uk} \\
\And
R. K. Mantiuk\\
  Computer Laboratory\\
 University of Cambridge\\
  Cambridge, United Kingdom \\
  \texttt{rkm38@cam.ac.uk} \\
     }

     \DeclareMathOperator*{\argmax}{arg\,max}
\DeclareMathOperator*{\argmin}{arg\,min}
     \newcommand{\code}[1]{\emph{#1}}

\begin{document}

\maketitle

\begin{abstract}

Most popular strategies to capture subjective judgments from humans involve the construction of a unidimensional relative measurement scale, representing order preferences or judgments about a set of objects or conditions. 
This information is generally captured by means of direct scoring, either in the form of a Likert or cardinal scale, or by comparative judgments in pairs or sets. 
 In this sense, the use of pairwise comparisons is becoming increasingly popular because of the simplicity of this experimental procedure. However, this strategy requires non-trivial data analysis to aggregate the comparison ranks into a quality scale and analyse the results, in order to take full advantage of the collected data. This paper explains the process of translating pairwise comparison data into a measurement scale, discusses the benefits and limitations of such scaling methods and introduces a publicly available software in Matlab. We improve on existing scaling methods by introducing outlier analysis, providing methods for computing confidence intervals and statistical testing and introducing a prior, which reduces estimation error when the number of observers is low. Most of our examples focus on image quality assessment.
%-------------------------------------------------------------------------
%  ACM CCS 1998
%  (see http://www.acm.org/about/class/1998)
% \begin{classification} % according to http:http://www.acm.org/about/class/1998
% \CCScat{Computer Graphics}{I.3.3}{Picture/Image Generation}{Line and curve generation}
% \end{classification}
%-------------------------------------------------------------------------
%  ACM CCS 2012
  % (see http://www.acm.org/about/class/class/2012)
%The tool at \url{http://dl.acm.org/ccs.cfm} can be used to generate
% CCS codes.
%Example:
% <concept_id>10010147.10010371.10010352.10010381</concept_id>
% <concept_desc>Computing methodologies~Collision detection</concept_desc>
% <concept_significance>300</concept_significance>
% <concept_id>10010583.10010588.10010559</concept_id>
% <concept_desc>Hardware~Sensors and actuators</concept_desc>
% <concept_significance>300</concept_significance>
% <concept_id>10010583.10010584.10010587</concept_id>
% <concept_desc>Hardware~PCB design and layout</concept_desc>
% <concept_significance>100</concept_significance>
\end{abstract}

\section{Introduction}

%Pairwise comparison experiments consider the comparison of objects in couples. 

%Aggregating ranks in the form of comparisons in a pairwise or setwise fashion is popular in applications that rely on non-expert evaluators, primarily when conditions can only be judged subjectively, e.g. taste testing, color comparisons or quality of experience.

\begin{figure}[t]
\centering
\includegraphics[width=0.75\textwidth]{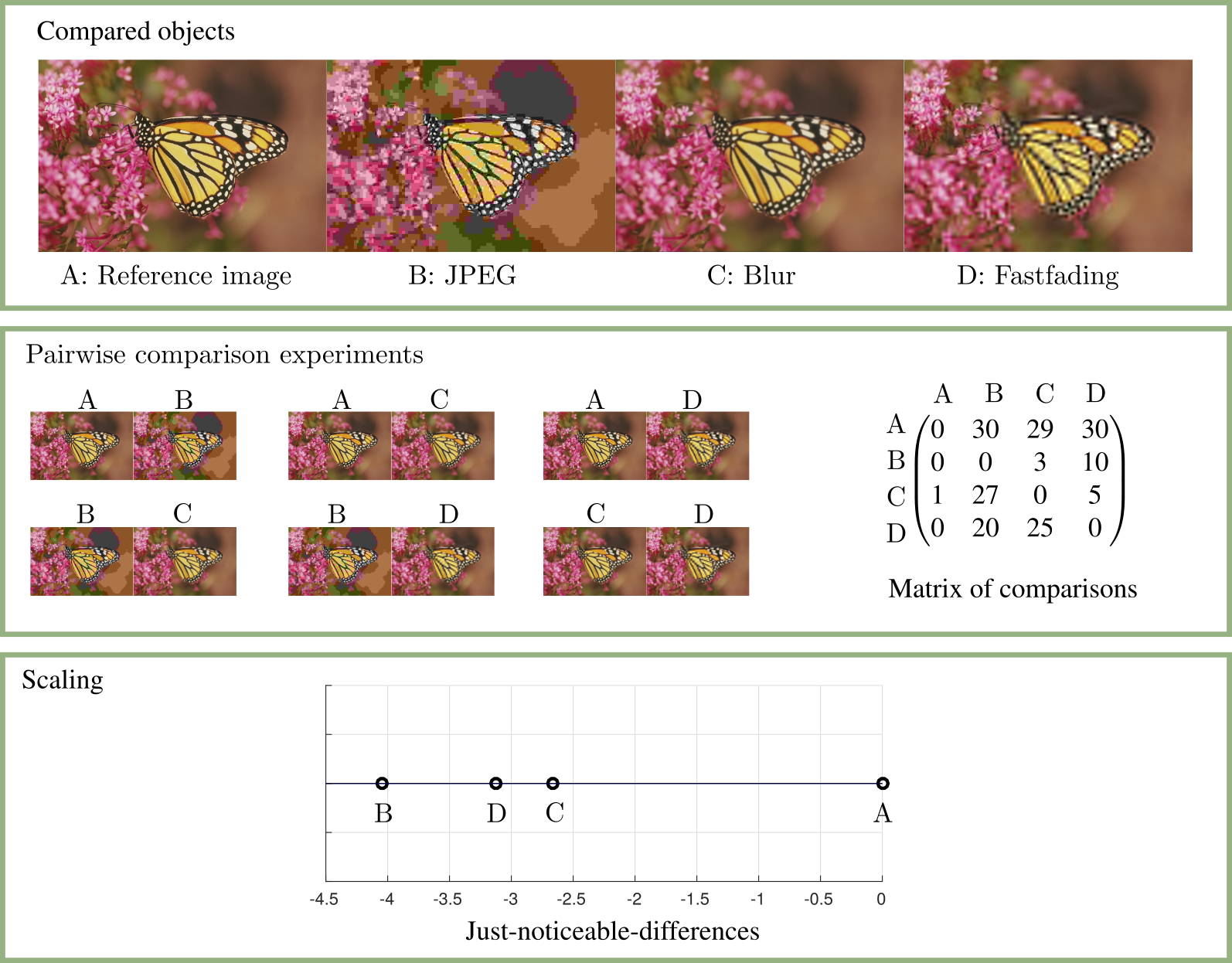}
\caption{Illustration of scaling pairwise comparison data for evaluating the perceived image quality. $4$ conditions (distortion types) are compared in this case, resulting in $\binom{5}{2}=5$ different comparisons, each comparison repeated $30$ times. Scaling algorithms produce a quality scale from the matrix of comparisons, in which distances between conditions can be interpreted as probability of better perceived quality.}
\label{fig:graph-abstract}
\end{figure}

One way to measure a perceptual attribute of interest, such as image quality, is to ask experiment participants to rank a set of conditions, for example images. The simplest type of such ranking are pairwise comparisons, where only two conditions are shown at a time and a participant is asked to choose one of them according to some specific criteria. For example, if we want to analyse which of three rendering methods (A, B and C) produces the highest quality results, we could present the images produced by these methods in pairs (AB, BC, AC) and then ask observers which image in each pair has better quality. 
If enough data is collected, we can then rank the algorithms from the best to the worst, estimate the confidence in such ranking, and scale the ranking scores so they can be easily interpreted in terms of probability of better perceived quality. A representation of this strategy can be seen in Figure \ref{fig:graph-abstract}. Unidimensional scaling methods attempt to represent preference judgments on a line, so as to effectively retain the distance information between the tested objects.
 This projection may reveal the underlying structure or unique relationships among the objects, allowing to measure and compare them in a meaningful way.

Pairwise comparison experiments are simple to run, but the data analysis step becomes more difficult.
%RM: Not essential info
%The results can be easily interpreted when only two conditions are compared, based on the number of votes. However, if more conditions are considered, the interpretation is more complex.
Often, data analysis is limited to statistical testing: showing that observed differences are unlikely to be produced by chance. Although this is an important stage of data analysis, it is often insufficient, as statistical significance may not translate into practical significance.
%If one method is preferred in 51\% of the cases, it is still possible to prove statistical significance. However, for most applications the difference between 51\% and 49\% is obviously not very relevant in practical terms. 
The scaling methods presented in this paper can express the results in terms of practical difference: they translate raw comparison data into quality scores that show the magnitude of the difference between tested conditions. %(i.e. they produce a rank of the objects compared, which can be easily interpreted).

% and confidence in the results.

There is a vast amount of literature on scaling or aggregating comparative judgments through pairwise comparison experiments, dating as early as 1927 \citep{Thurstone1927,Davidson1976}. However, studying this literature could be a daunting and time consuming task, which requires a strong background in statistics to understand all the intricacies of these methods. 
 Scaling methods are usually not straightforward to implement. They require a number of precautions to ensure robust results and that errors are not introduced due to insufficient floating point precision and other non-obvious reasons. One of the purposes of this paper is to provide a comprehensive description of how scaling methods work, and accompany this with an open source Matlab toolbox for performing the scaling and statistical analysis. %and for simulating experiments. %The availability of free software for scaling is usually limited to specialised statistical packages and requires good knowledge of those to use. %Furthermore, the existing software can rarely estimate confidence intervals, which are provided by our toolbox. 

%This paper is focused on data analysis, however, we point out a few %key aspects of experiment design in Section~\ref{sec:exp-design}. 

% The novel contributions of this work are the following: (i) an improvement in terms of precision of a state-of-the-art scaling procedure for small sample sizes (achieved by adding a prior), (ii) a novel outlier detection technique, which highlights observers that do not behave similarly to the rest of the sample, (iii) the computation of confidence intervals, and (iv) insight on several practical issues concerning the experimental design. 
% %The prior reduces the size of confidence intervals as well as the estimation bias if data is collected from less than 20 observers. The outlier detection method helps to identify cases where the answers from an observer greatly differ from the rest of participants. 

% RM: We cannot have this for anonimity reasons
The scaling method described here has been used in several previous computer graphic projects of our and other groups, including \citep{Karaduzovic-Hadziabdic2016,Eilertsen2015,Vangorp2014,Wanat2014}.
However, due to space restrictions, we could not explain in those papers all the details and improvements. This paper is meant to serve as a reference for any future work relying on our scaling method. 
%\textcolor{red}{What makes the paper focused on image and video quality?}

The contributions of this work are the following:
(i) a collection of methods for the analysis of pairwise comparison data, which include outlier analysis, estimation of confidence intervals and statistical testing; (ii) a prior, which improves scaling accuracy when the number of observers is low; (iii) analysis of practical issues concerning the experimental design, such as the use of ties or incomplete designs; and (iv) a Matlab toolbox to perform the analysis.
%The experiments performed in this paper consider different simulations and real-world image and video quality datasets.  We use mainly simulated experiments because only in this case we know what the true values of the quality scores are.

\subsection{Direct rating vs. pairwise comparisons}
Direct rating, in which observers assign a score to each condition, may seem to be a simpler and more direct measurement of perceptual attributes (e.g. image quality or taste) than pairwise comparisons. However, direct rating methods have a number of limitations. They require careful training so that participants know what value should be assigned to which condition --- to establish a well defined scale for a given experiment. However, even after careful training, such scale can vary substantially between participants, or even within a single participant when the experiment is repeated on different days. Direct rating experiments are particularly difficult to conduct when compared conditions are substantially different from each other. For example, the popular LIVE image quality dataset \citep{Sheikh2006b} was collected in 7 different experimental sessions, where each session involved only one type of distortion (e.g. JPEG compression, noise, blur, etc.). Isolating each distortion type simplified the experimental task, but it made the quality scale obtained in each session different from one another. To align all scales, the authors had to perform 8 realignment experiments, in which a subset of images from the 7 experimental sessions was assessed again and the collected scores were used to linearly re-scale the previously collected scores. This rather complex procedure demonstrates the challenges of obtaining a unified quality scale in rating experiments.

As opposed to this, the use of pairwise comparison present numerous advantages: i) it leads to a very simple experimental task and is therefore well suited for non-expert participants, ii) it avoids calibration issues frequently encountered in cardinal measurements \citep{Tsukida2011}, iii) it generally provides higher sensitivity and a lower measurement error when compared to direct rating \citep{Shah2015}, and iv) it can be faster to run than direct scaling (particularly since making pairwise comparisons is  easier and faster for participants \citep{Stewart2005} and because the number of comparisons can be reduced using adaptive procedures \citep{Mantiuk2012a,Ye2014,Xu2011}).

\subsection{Vote counts vs. scaling}
The simplest way to report the result of a pairwise comparison experiment is to compute vote counts --- the number of times one condition was selected as better than any other condition. Vote counts, however, present the results on an ordinal scale, which would usually produce the correct ranking of the conditions, but it does not correctly capture the magnitude of the differences between conditions. On the other hand, pairwise comparison scaling places those conditions on a continuous interval scale, which captures both the order of conditions and the magnitude of the difference. Zerman et al. compared the results of pairwise comparison scaling and vote counts to the scores obtained in a direct rating experiment \citep{zerman2017}. They showed that scaled data is more strongly related to rating scores than vote counts, confirming that quality magnitudes are better captured when pairwise comparison data is scaled. Furthermore, vote counting is difficult when not all conditions are compared with each other (incomplete design) or when not all observers compare the same conditions (unbalanced design). Scaling methods can robustly cope with such non-standard experiment designs. 

\section{Related work}
\label{sec:relatedwork}

% http://www.sciencedirect.com/science/article/pii/S030505480300042X
% http://www.sciencedirect.com/science/article/pii/S0377221796002500

The bibliography in papers that review how to aggregate pairwise comparison data testifies the widespread interest of the scientific community on this type of methods: more than 350 papers in \citep{Davidson1976} and more than 100 in \citep{Cattelan2012}. There is a wide range of applications in which this approach has shown to be successful, e.g. to study consumer preference, in sport rankings, econometrics or perceived image/video quality. For a detailed discussion on the topic the reader should refer to one of the aforementioned review papers or to the monograph of David \citep{David1963}. An accessible introduction to the topic of scaling can be found in \citep{Tsukida2011} and \citep{Dunn2004}.

The scaling procedure depends on the selection of the model relating observers' answers to the abstract quality scale. Two of the most common models are that of Thurstone \citep{Thurstone1927} (considered in this work) and of Bradley and Terry \citep{Bradley1952}. The differences between the two models are minor \citep{Tsukida2011} and the choice is a matter of preference. Many extensions of the two models can be found in the literature. For example, some models give the observers an additional option of choosing tie (no preference) \citep{Davidson1970} or let them express strong, mild, or no preference judgment for a pair conditions \citep{Agresti1992}. While the answers are typically considered to be independent, some literature focuses on learning from dependent data \citep{Cattelan2012}, where either condition and observer covariates are accounted for. Condition dependencies stem from the fact that generally the same condition is involved in multiple paired comparisons. Modeling observer covariates assumes that the comparisons made by the same person are dependent. Allowing ties or accounting for covariates is not free from shortcomings. Those more complex models usually require more data as more parameters need to be estimated. Other type of models introduce a temporal component \citep{Herbrich2006}, e.g. for ranking tournament data, where players take part in different matches during a prolonged period of time and most recently played matches need to have more influence on the ranking to account for changes in the skills of different players.

It is also worth noting that pairwise comparison experiments are not only used for measuring stimuli on interval scales, as presented in this paper, but can also be used to discover unknown perceptual attributes. This is, they can be used to discover explanatory variables that affect the results of the comparisons \citep{springall1973response}. This can be performed by standard statistical approaches (e.g. regression analysis), multi-dimensional scaling \citep{Pellacini2000} or using more advanced ranking machine learning methods \citep{Wauthier:2013:ERP:3042817.3042949}. 
Paired comparisons are usually expected to be consistent, which may not hold in practice. In some cases, preferences can be naturally intransitive (i.e. A>B, B>C but C>A), which usually originates from the fact that the conditions have more than one aspect of interest, and different aspects prevail in different comparisons. As said, some approaches account for this \citep{Causeur2005a,usami2010}, and project the data to more than one dimension by multi-dimensional scaling. This approach might simplify the task of scaling but makes more difficult the interpretation of the final solution.  

The pairwise comparison scaling method discussed in this paper is only suitable when the quality differences between compared conditions are small so that the observers vary in their answers. When a perceptual attribute must be scaled over a larger range, the difference scaling method \citep{Maloney2003} could be more appropriate. In this method observers are asked to judge the magnitude of a difference for two pairs of stimuli and select the pair of higher difference.

%Bayesian Network (TrueSkill): 

Most of the software facilitated to work with paired comparison data is implemented in R, where apart from most traditional models, one can also find more specific techniques.
%Here we review some of the code that can be found for this purpose and give a brief overview of the main strengths of each toolbox.
The \emph{eba} package \citep{Wickelmaier2004} adapts one of the most popular models (the Bradley-Terry model) to consider that different conditions might present various aspects that account for their worth (referred to as elimination-by-aspects models) and includes different functions to check the consistency of the answers (i.e. violations of transitivity). 
The \emph{prefmod} package \citep{Hatzinger2010} implements also different versions of the Bradley-Terry model. No preference options (ties) can be included and specifically modelled, as well as incomplete designs. 
The \emph{Bradley-Terry2} package \citep{Turner2012} includes different probability functions and models data covariates. This package also allows the use of tournament data. 
%The software in psychotree \citep{Strobl2011} mainly focuses on preference trees, that partition observers in a tree and fit a model to each subgroup. 
The package \emph{choix} in python presents inference algorithms based on an extension of Bradley-Terry model \citep{Placket1975}, which allow to explain and model comparisons between items, not only in a pairwise manner, but also setwise \citep{Maystre2015}. 
Finally, the tutorial in \citep{Tsukida2011} also includes some basic scaling code in Matlab. Although the mentioned software serves a similar purpose as our proposed method, none of the packages offers a complete set of methods for analysis, including outlier analysis, the computation of confidence intervals and statistical testing. 

%Some of the methods included in these packages are equivalent to our formulation, but we include some other novel features, as referred in the introduction.

This work is inspired by the previously mentioned papers, however, our focus is on more practical issues of scaling, such as experimental design, statistical analysis and low sample scenarios, providing guidance for the end-user of this type of methods and software for performing the scaling and analysis of results. 

% \textcolor{red}{Do we want to compare to any of these methods?}
% 
% \textcolor{red}{The paper in \citep{Wickelmaier2004} mentions that for the BTL model, Bradley (1955) has described how
% to estimate confidence intervals for the MLEs based on the Hessian and covariance matrices. Their code, however, (eba package) includes a function to perform bootstrapping for computing confidence intervals, but apparently it is mainly used to do the elimination-by-aspects part and see the confidence interval of each aspect. They do not mention any of this in the paper though. }

%Tutorial: Different models in matlab. Slow optimisation, weak experiments.
 
%  Include here some paragraph describing the existing code as well \citep{Cattelan2012}. R: prefmod, eba, BradleyTerry2. Matlab: tutorial. 
% \textcolor{red}{Check if other packages include CI (yes) and outlier analysis.}

%=========================================================
\section{Example of pairwise comparison data analysis}
\label{sec:software}
\label{sec:example-scaling}

We start by presening an example\footnote{The code for the example can be found in the 
\code{examples} folder, under the name of \emph{video\_TMO\_analysis\_example}.} of a typical pairwise comparison data analysis session using our software\footnote{\url{https://github.com/mantiuk/pwcmp}}, in which we analyse the data from the video tone mapping evaluation project presented in \citep{Eilertsen2013}. 

\begin{table*}[h!]
\small
\centering
\begin{tabular}{ r | r | r | r| r |r }
Observer & Session & Scene & Condition\_1 & Condition\_2 & Selection \\ \hline
1 & 1 & Window	& TMO\_Camera	& Ferwerda96 & 1 \\
1 & 1 & Exhibition & Ronan12 & Irawan05	& 2 \\
1 & 1 & Corridor & Irawan05 & Ferwerda96 &	1 \\
2 & 2 & Corridor &	Ronan12 & TMO\_Camera & 2\\
\end{tabular}
\caption{Formatting example recommended to store pairwise comparison data.}
\label{tab:csvexample}
\end{table*}

We recommend to keep the data in a tabulated format, such as comma-separated-files (CSV), in which each condition is described by meaningful labels. Such files are easy to read with any software and can be easily interpreted even long after the data have been collected. Table \ref{tab:csvexample} shows a few rows from the analysed dataset.

The first step is to convert the answers from the table into a set of comparison matrices ${M}$, one matrix per each observer. In such a matrix, columns and rows correspond to compared conditions and matrix value $c_{ij}=n$ means that condition $O_i$ was $n$ times selected as better than condition $O_j$. If there is a reference condition, such as a non-distorted image, it should be put in the matrix as the first condition in the first row and column. The first condition will be assigned a fixed quality value of 0. 
%It may seem easier to store experimental results directly as comparison matrix, however, storing the data as proposed here allows a more broad data analysis (e.g. analysis per observer or scene). 

\begin{figure}[t]
\centering
\includegraphics[width=0.7\textwidth]{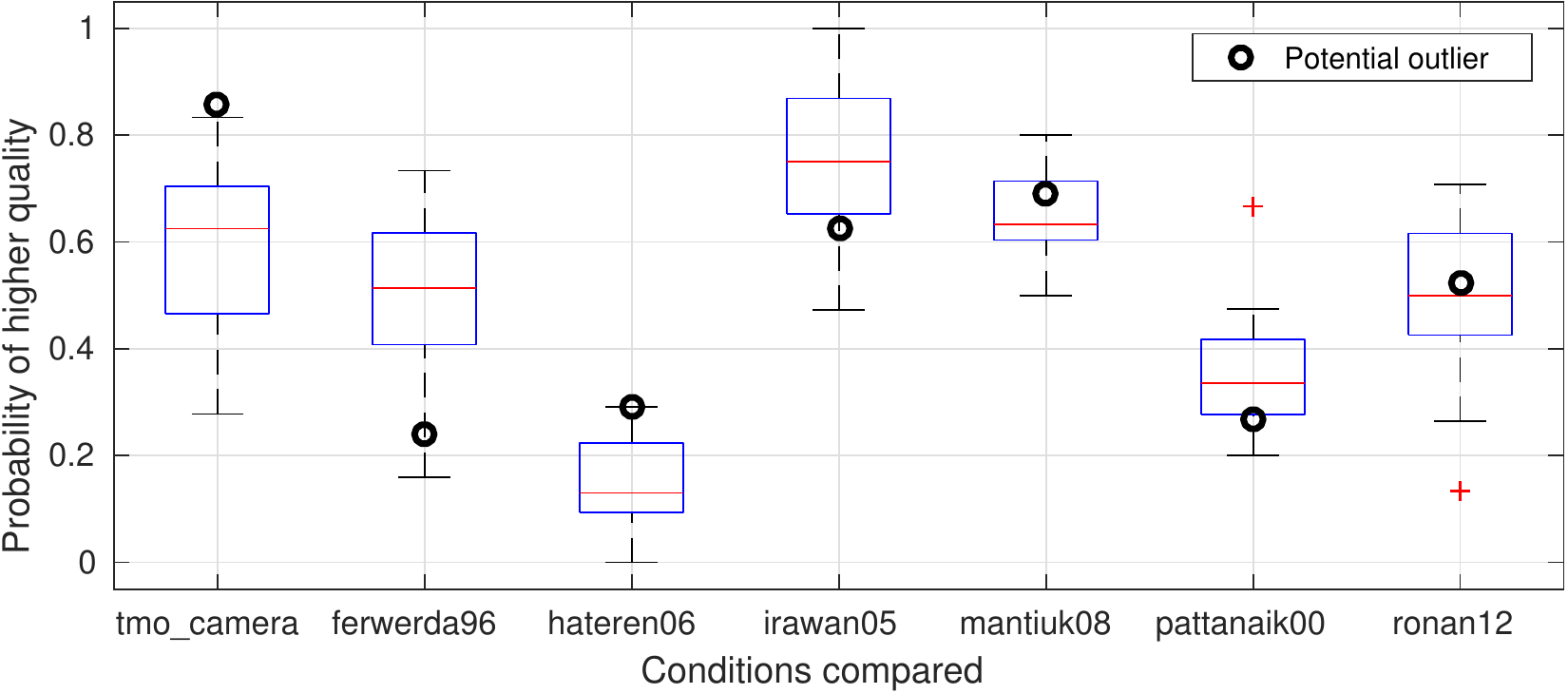}
\caption{Distribution of the general perceived quality for each condition. This plot should be interpreted with care and never be used as a replacement of the scaling. In this case, the probabilities plotted here and the scaling in Figure \ref{fig:errorbars} show similar behaviour because the dataset uses full design, but they would be very different for incomplete designs.}
\label{fig:outlier-boxplot}
\end{figure}

The second step is to perform outlier analysis to detect potential observers who performed very differently from the rest. The function to perform this analysis is \code{[L,L\_dist]=pw\_outlier\_analysis(M)}, which receives a matrix $M$ with the responses per observer and returns the likelihood $L$ of observing the data of each observer and a inter-quartile-normalised score $L\_dist$, which indicates the observers that should be further investigated. Since there is no objective threshold that could distinguish outliers with high confidence, we advise to investigate all observers whose $L\_dist$ score is close or above the customary threshold of 1.5.
%\textcolor{red}{I am not sure if it's a proper way of showing the sintax of the functions. I think we may need to decide whether we want to go for a 13 or 14 page paper and reduce some parts in that case. }
%The results for the $18$ observers in the analysed dataset are the following: $(2.72, 0, 0, 0.98, 0, 0, 0, 0, 0, 0, 1.37, 0, 0, 0, 0.15, 0, 0, 0)$, 
The results for the $18$ observers in the analysed dataset indicate that there is one observer with a score of 2.72, which requires further attention.
%There are other two observers (with scores of 1.37 and 0.98) that can be further analysed as well, but this is optional, as the threshold is usually set up at 1.5.
To compare the answers of the indicated observer (observer number n\_obs) to the rest of observers, we use the function \emph{compare\_probs\_observer(M,n\_obs)}, which plots the probabilities of selecting one condition over all others, shown in Figure~\ref{fig:outlier-boxplot}. Note that this presentation of the data does not involve scaling, which could obscure the patterns that are specific to an outlier. 
%To analyse the difference between the potential outlier and the rest of the observers (which we strongly recommend in order not to remove valid observers), see Figure~\ref{fig:outlier-boxplot}. The figure shows the probability of each condition being considered as better than the rest. This is represented for all of our observers with a boxplot (summing up all the results for the 5 scenes of this dataset).
The black circles in the plot represent the answers of the potential outlier.
%The plot also shows the results obtained by an observer that would perform completely at random (resulting in an inter-quartile-normalized score of 10.38).
The plot indicates that the potential outlier had a different opinion about operators \emph{Ferwerda96}, \emph{Hateren06} and \emph{TMO\_Camera}, but the patterns of his answers were not much different from the rest of observers. Therefore, although the observer was not fully consistent with the rest of observers, we could not justify removing her/his answers from the dataset. We recommend performing such detailed per-observer analysis, rather than using arbitrary measure to exclude observers. 
%, it can be seen that the perceived quality for  is slightly different than the rest. (\textcolor{red}{Link some information of the dataset to this}). Because of this, we believe that this result is not suspicious and decide not to remove this observer from the dataset.
%In this case, the outlier seems to have a lower perceived quality when compared to the rest for Ferwerda96, Irawan05 and Pattanaik00. However, the observer perceives Hateren06 and TMO\_Camera as cases of much greater quality. It is noticeable that the observer perceives Hateren06 as of greater quality than Ferwerda06, when this is not the case for the rest of the sample (the same applies to Irawan05 and TMO\_Camera). As a result, we decide to remove this observer for the subsequent analysis. 
%Note that this analysis can be also performed independently per scene.
The details of the outlier analysis can be found in Section~\ref{sec:outlier}.

\begin{figure}[t]
\centering
\includegraphics[width=0.7\textwidth]{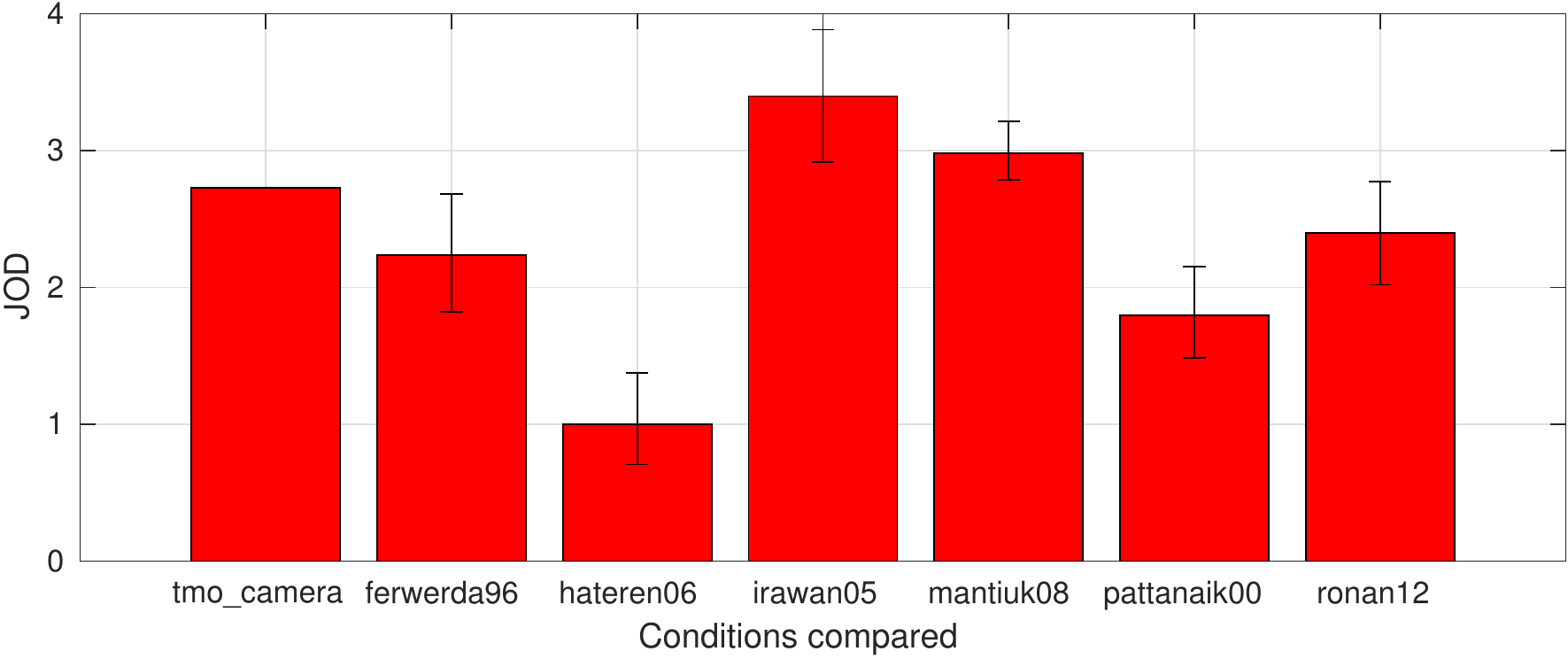}
\caption{Visualization of the scaling results and confidence intervals for the chosen dataset. Note that there is no confidence interval for the first condition, as this is always set up to a fixed value (since scores are relative). The difference of 1 JOD indicates that 75\% of observers selected one condition as better than the other. }
\label{fig:errorbars}
\end{figure}

\begin{figure}[t]
\centering
\includegraphics[width=0.7\textwidth]{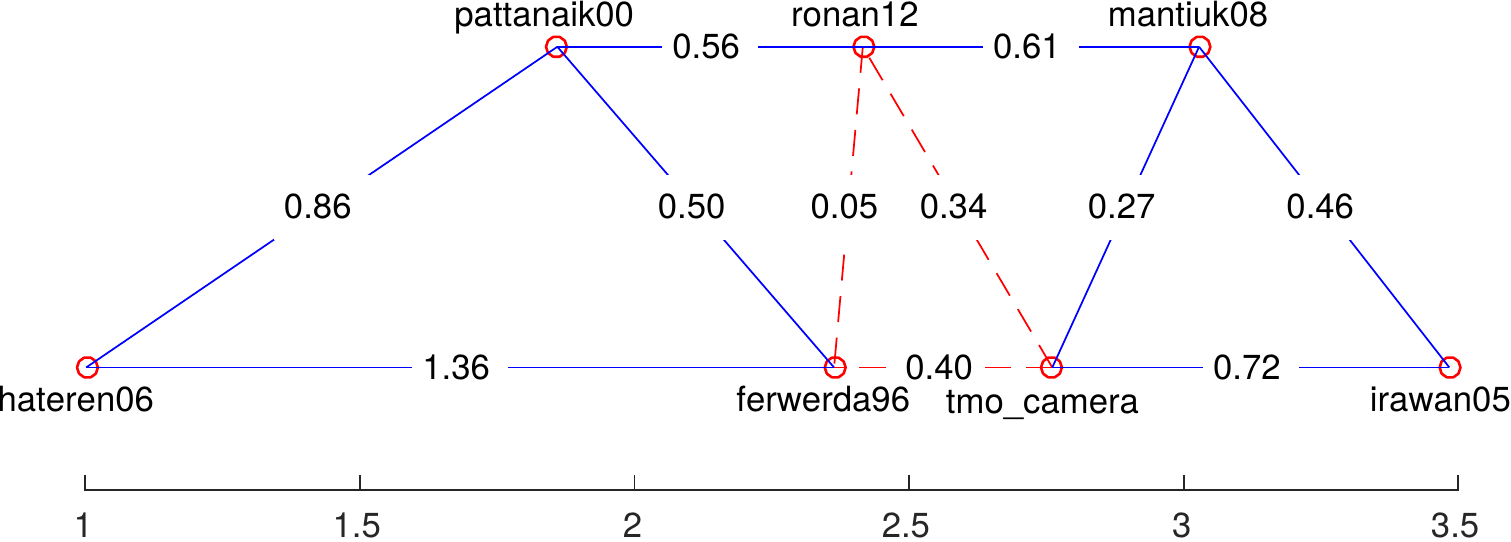}
\caption{Graphical representation of the scaling. Red points represent conditions, and they are only connected to their neighbours, as these are usually the comparisons in which we are most interested. Blue solid lines represent statistically significant differences, as opposed to red dashed lines. The x-axis shows the scaling.}
\label{fig:triangles}
\end{figure}

Once we are confident there are no outliers in the dataset, we can scale the results and compute confidence intervals using \emph{[jod, stats]=pw\_scale\_bootstrp(M)} function. The function expects the same matrix of comparison per observer $M$ as the outlier analysis and returns the scaling solution and a set of statistics. The scaling and the confidence intervals have been plotted for our dataset in Figure~\ref{fig:errorbars}. Confidence intervals represent the range in which the estimated quality values lie with 95\% confidence. The confidence intervals, however, should not be used to infer statistical significance of the difference. The statistical tests are performed by the function \code{pw\_plot\_ranking\_triangles(jod,stats)}, which produces a plot shown in Figure~\ref{fig:triangles}. The continuous lines in that plot indicate statistically significant difference between the pair of conditions and the dashed lines indicate the lack of evidence for statistically significant difference. More information on the scaling and statistical analysis can be found in Sections~\ref{sec:scaling} and \ref{sec:stat-analysis}, respectively. 

%If in need of a more simple and light approach in computational terms, the \emph{[jod]=pw\_scale(C)} function can be used, which receives the matrix of comparison $C$ summed up for all observers and returns the scaling solution without performing the statistical analysis. For a more detailed explanation of the methods presented here and pairwise comparison experiments please refer to: Section~\ref{sec:scaling} for the scaling method, Section~\ref{sec:prior} for the distance prior proposed and
%Section~\ref{sec:practical} for some of the practical issues that arise when performing pairwise comparison experiments. 

% The scaling can be done by means of the \emph{pw\_scale} function, which receives the complete matrix of comparisons. The scaling obtained in this case for the $5$ considered conditions is: $(0, -1.3840, 1.1414, 0.6689, -0.4906, 0.0940, 0.3693)$, meaning that the condition of lowest quality is Hateren06 and the one with the highest quality is Irawan05. The scaling considering also the outlier is: $(0, -1.2378, 1.1609, 0.7416, -0.4425, 0.1611, 0.4920)$, where it can be seen that the ranking of the conditions is maintained, but the scaling is different. 

\section{Designing pairwise comparison experiments}

Planning for pairwise comparison experiments requires taking into account several considerations to ensure that sufficient data is collected with possibly small experimental effort. The number of required comparisons depends on the number of compared conditions $n$ (e.g. different algorithms or distortion levels), the number of different pieces of content $k$ (e.g. images or video clips) and the number of repetitions $t$ of the experiment. If each observer is asked to compare each condition with the rest, they would need to perform $\frac{1}{2}{\cdot}n{\cdot}(n-1){\cdot}k\cdot t$ comparisons. This number grows quickly, especially for large $n$.

%  Generally, having observers that perform different comparisons is non-advisable \citep{Cattelan2012}. However, the number of needed comparisons can be significantly reduced by the use of more intelligent approaches \citep{Ye2014,Xu2011}, minimising experimental effort while keeping a competitive performance. Most of the work in this sense revolves around the idea of ordering objects to compare neighbors in the quality scale \citep{Xu2011,Silverstein2001}. This is because these are the most meaningful comparisons, provided that differences are large enough for observers to distinguish between these conditions. 
%  More recent approaches explore the use of active sampling through information gain metrics \citep{Ye2014,conf/nips/JamiesonN11}.

An important issue is the choice of compared conditions, since not all comparisons are equally useful.
The comparisons that produce obvious results, e.g. comparing the highest and lowest distortion levels, do not contribute much to the outcome of the experiment and can be obviated. The experiments, in which only selected pairs are compared, are referred to as incomplete design, as opposed to a complete design, in which every pair is compared.
If not all observers compare the same set of conditions, but instead, every observer has a different experimental design, the experiment is said to have an imbalanced design. Note that this imbalanced design is generally non-advisable \citep{Cattelan2012}, at least when not accounting for observer covariates. 
If we do not have any apriori information about the potential ordering of our conditions, we could use an efficient sorting algorithm (e.g. quicksort) \citep{Maystre2017} or other specifically designed techniques, such as active sampling \citep{Ye2014,conf/nips/JamiesonN11}. This results in less variance given the same number of trials \citep{Silverstein2001,Shah2015}. %The disadvantage of using sorting algorithms is that the presentation of the pairs is not randomised, which may lead to bias due to the learning effect (observer getting familiar with the content and, for example, becoming better at spotting distortions). 
In many cases, however, we know in advance the most likely order of the conditions, e.g. in image compression we know that lower bit-rate images will have worse quality than those of higher bit-rate. In such cases, we can restrict comparisons to neighbours in the scale of distortion level. It is important to ensure that the quality levels of compared images are relatively similar, so that they are confused in certain number of cases. If all observers give the same response, we will not be able to reliably estimate the scaled difference between them.
This is further discussed in Sections~\ref{sec:prior} and \ref{sec:fullvsincomplete}.

Finally, it is possible to offer a third answer in the experiment (i.e. ties). 
This, however, usually makes modelling more difficult. We discuss this problem in more detail in Section~\ref{sec:ties}. Our general recommendation is to run two-alternative-force-choice experiments without ties.

Discussion of other important factors related to experimental design, such as control of the viewing conditions, reducing learning effects, training, experimental fatigue, are out of the scope of this report. Readers can refer to \citep{Engeldrum} or psychophysics textbooks, such as \citep{Kingdom2016} or \citep{Lu2013}. 

\section{Problem formulation}
\label{sec:scaling}

Suppose we aim to compare $n$ conditions $O_1,\ldots,O_n$ (e.g.
$n$ images, generally with the same content, each processed with a different algorithm) with unknown underlying true quality scores $q=(q_1,\ldots,q_n), q_i \in \mathbb{R}$. 
%Most traditional models assume only two possible outcomes of each comparison, so that the random variable associated to each paired comparison is a binary random variable.
% The probability of $q_i>q_j$ (denoted as $p_{ij}$), will directly depend on the worth of conditions $O_i$ and $O_j$: 
% \begin{equation}\label{eq:distributionf}
%  p_{ij} = F(q_i- q_j),
% \end{equation}
% $F$ being the cumulative distribution function of a random variable. 
The aim of this analysis is to estimate scores $\hat{q}=(\hat{q}_1,\ldots,\hat{q}_n)$ that approximate the true quality scores $q$. This can be
obtained from the pairwise comparisons collected from $m$ observers in $t$ trials (and possibly $k$ pieces of content, each processed separately). Because the pairwise comparisons are relative, we also assume that $q_1=\hat{q}_1=0$.

\subsection{Comparison matrix}

A pairwise comparison experiment is usually represented in a count matrix $\mathbf{C}$, where each element $c_{ij}$ measures the number of cases in which condition  $O_i$ has been selected as better than condition $O_j$ (considering $m$ observers and $t$ trials). For example, in an experiment with  three conditions, the resulting matrix could be as follows:
\begin{equation}
\mathbf{C}=\begin{bmatrix}
     0  &   3  &   0 \\
    27  &   0  &   7 \\
    30  &  23  &   0	
\end{bmatrix}
\label{eq:measurement-matrix}
\end{equation}
%where each column and row of that matrix corresponds to one condition. 
This is, $c_{12}=3$ tells us that condition $O_1$ has been selected three times as being better than condition $O_2$, and $c_{21}=27$ tells us that condition $O_2$ has been selected 27 times as better than $O_1$. The probability that one condition is selected as better than another (denoted as $p_{ij}$ for $O_i$ and $O_j$) can be estimated using the empirical information in matrix $\mathbf{C}$ \citep{Tsukida2011}: 
\begin{equation}\label{eq:probfromC}
 \hat{p}_{ij} = \frac{c_{ij}}{c_{ij}+c_{ji}}, \;\; i\neq j
\end{equation}
e.g. the probability that $O_2$ is selected as better than $O_1$ can be estimated as $\frac{c_{21}}{c_{21} + c_{12}}=\frac{27}{27+3}=0.9$.

% which give us: 
% \begin{equation}
% \mathbf{P}=\begin{bmatrix}
%      -  &   0.033  &   0.000 \\
%     0.967  &   -  &   0.233 \\
%     1.000  &  0.767  &   -	
% \end{bmatrix}.
% \label{eq:matrixprobabilities}
% \end{equation}

\subsection{Observer model}

% The probability of $q_i>q_j$ (denoted as $p_{ij}$), will directly depend on the worth of conditions $O_i$ and $O_j$: 
%  \begin{equation}\label{eq:distributionf}
%   p_{ij} = F(q_i- q_j),
%  \end{equation}
%  $F$ being the cumulative distribution function of a random variable.

% [This is repetition as we have discussed it in the related work]
%Most scaling methods for paired comparisons assume a specific distribution: Thurstone's Case V model \citep{Thurstone1927}, which assumes that the difference between the quality scores follows a Gaussian distribution, or Bradley-Terry model \citep{Bradley1952}, which assumes that such a difference is explained by a logistic distribution. A discussion of both models can be found in \citep{Tsukida2011}. In practice both models produce very similar results. %We will use Thurstone's Case V model in this paper. 

\begin{figure}[t]
\centering
\includegraphics[width=0.6\textwidth]{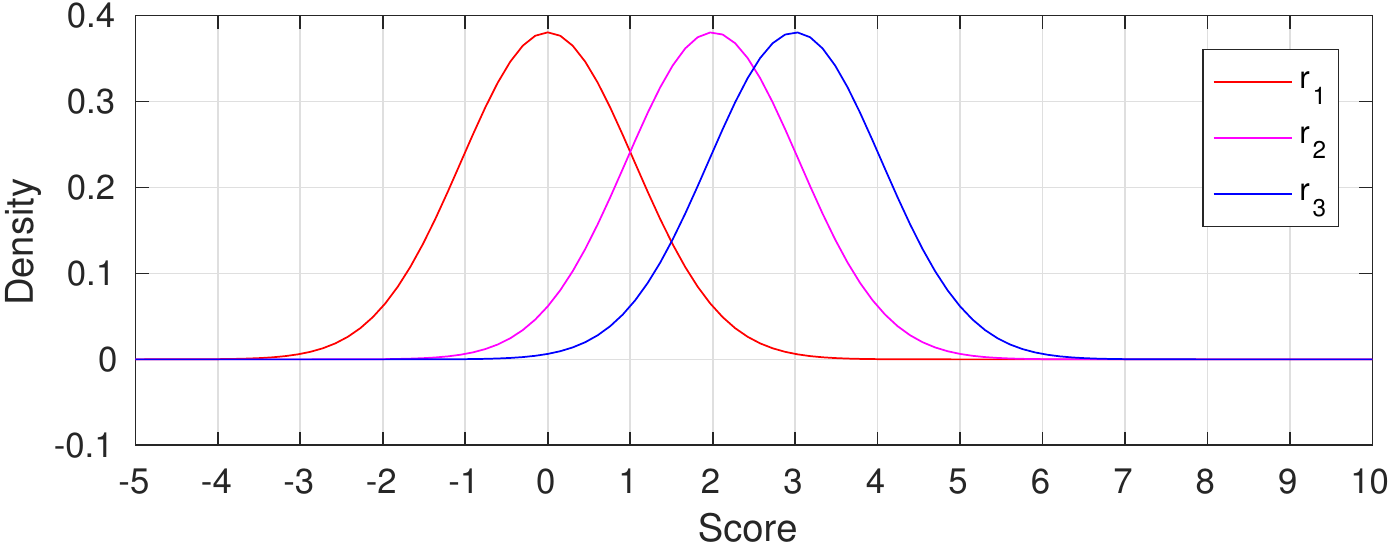}
\caption{The probability of assigning a score to three conditions. Quality scores are assumed to be random variables. }
\label{fig:three-quality-scores}
\end{figure}

%As stated before, scaling pairwise comparison data usually requires the assumption of a certain underlying model which describes how observers assess image quality. 
%The most common one is the 
In this paper we use the model proposed by L. L. Thurstone \citep{Thurstone1927,Engeldrum}. 
This model assumes that observers make quality judgements by assigning a single quality value to each condition and that the condition's quality is a random variable, so as to account for the subjective nature of these experiments. This is, the perceived quality of a condition $O_i$ is modeled as a random variable: {$r_i \sim N( q_i,\sigma)$ (i.e. the mean of the distribution is assumed to be the true quality score $q_i$). This is illustrated on an example of three conditions in Figure~\ref{fig:three-quality-scores}.
%The observers on average judge condition C to have higher score than B, but it is also probable that an observer will make the opposite judgement. 
Observers vary in their notions of quality among them (inter-observer variance), and their opinions are also likely to change when they repeat the same experiment (intra-observer variance). Thurstone Case V model assumes that both inter- and intra-observer variance can be explained by a Normal distribution, and that the variance of that distribution is the same for each condition (the noise parameter $\sigma$ is the same for all items and accounts for the uncertainty in the comparisons)}. The goal of the pairwise comparison experiment is to find the expected values $\hat{q}$ of the distribution of the scores for each condition.  In practice, since scores are relative, we are interested in recovering the distances among them.

\subsection{JNDs and JODs}

%The meaning and interpretation of the pairwise comparison data much %depends on the experimental question asked. The typical question in %such experiments is ``which of the two images [video clips] has  %higher quality / looks more natural / is less distorted?''. Given %such task, the observers choose an image that is better in some %respect, but they make this judgment relative to an imaginary perfect %quality reference image or video. Even if a reference content is not %directly presented to the observers, everyone has a good idea how an %image without distortions or looking more natural may look like. %Therefore the answer does not 

\begin{figure}[t]
\centering
\includegraphics[width=0.7\textwidth]{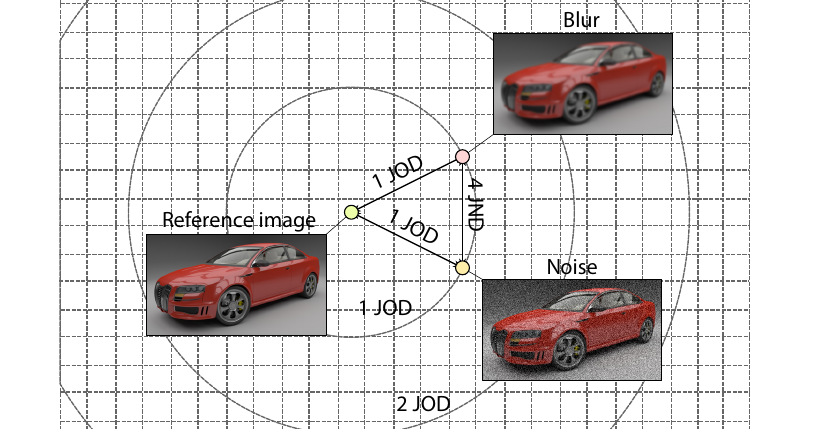}
\caption{ Illustration of the difference between just-objectionable-differences (JODs) and just-noticeable-differences (JNDs). The images affected by blur and noise may appear to be similarly degraded in comparison to the reference image (the same JOD), but they are noticeably different and therefore several JNDs apart. The mapping between JODs and JNDs can be very complex and the relation shown in this plot is just for illustrative purposes.}
\label{fig:jod_vs_jnd}
\end{figure}

The results of paired comparisons are typically scaled in Just-Noticeable-Difference (JND) units \citep{Engeldrum,Silverstein2001}. Two stimuli are 1 JND apart if 75\% of observers can see the difference between them. However, we believe that considering measured differences as ``noticeable'' leads to an incorrect interpretation of the experimental results.  
Let us take as an example the two distorted images shown in Figure~\ref{fig:jod_vs_jnd}: one image is distorted by noise, the other by blur. They are definitely noticeably different and intuitively they should be more than 1 JND apart. However, the question we ask in an image quality experiment is not whether they are different, but rather which one is closer to the perfect quality reference. Note that a reference image does not need to be shown to answer this question as we usually have a mental notion of how a high quality image should look like. Therefore, the data we collect is not related to visual differences between images, but rather to image quality difference in relation to a perfect quality reference. For that reason, we describe this quality measure as Just-Objectionable-Differences (JODs) rather than JNDs. Note that the measure of JOD is more similar to visual equivalence \citep{Ramanarayanan2007a} or to the quality expressed as a difference-mean-opinion-score rather than to JNDs.

\section{Scaling methods}

Pairwise comparisons can be viewed as noisy samples of the underlying quality difference between two conditions. The goal of scaling is to estimate these latent differences based on noisy data in the form of pairwise comparisons. 
Given the observer model, we can use one of the following methods to transform collected probabilities $\hat{p}_{ij}$ into scaled quality scores $\hat{q}$.

%Once the comparison matrix $\mathbf{C}$ has been collected, there are several methods that can be used for scaling the data.

\subsection{From probabilities to distances}

\begin{figure}[t]
\centering
\includegraphics[width=0.3\textwidth]{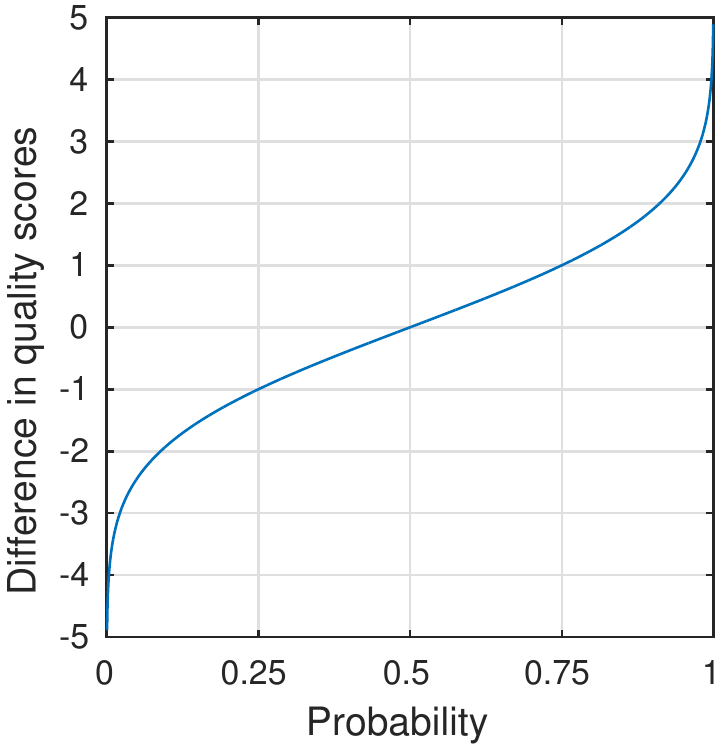}
\caption{Inverse cummulative Normal distribution mapping probabilities into distances on the JOD scale. When the observers make a random guess ($p_{ij}=0.5$), the JOD distance between conditions $i$ and $j$ is 0. When the observers select the first condition as better 75\% of the times ($p_{ij}=0.75$), the JOD distance is 1. The distance is negative when the observers  select the second condition more often as being better. }
\label{fig:cumulative-inv-norm}
\end{figure}

When scaling data, we are mostly interested in recovering the distance $q_i -q_j$ between underlying quality scores $q_i$ and $q_j$ (since scores are relative). This distance is linked to the probability of condition $O_i$ having a higher quality than condition $O_j$.
%Differences between scores can be computed from probabilities (which we estimated using \eqref{eq:probfromC}), assuming again a Normal distribution.
%This is because when we compare two quality scores to compute the probability of a condition $O_i$ having a higher quality than another condition $O_j$, $P(r_i>r_j) = P(r_i - r_j >0)$, we also assume a Gaussian distribution. 
Note that the difference of two Gaussians $r_i$ and $r_j$ is also a Gaussian random variable:
\begin{equation}
 r_i - r_j \sim N(q_{ij}, \sigma_{ij}),
\end{equation}
where $q_{ij} = q_i - q_j$ and $\sigma^2_{ij} = \sigma_i^2 + \sigma_j^2 = 2\sigma^2$.

%Note that the mean $\mu_{ij}$ (or expected value) corresponds to the difference in quality scores $q_i - q_j$. 

The probability of choosing $O_i$ over $O_j$ can be computed using the cumulative Normal distribution $\Phi$ over the difference $r_i - r_j$:
\begin{eqnarray}
P(r_i>r_j) &= P(r_i - r_j >0) = \Phi \left ( \frac{{q}_i-{q}_j}{\sigma_{ij}} \right )  \label{eq:prob-diff}
\\&= \frac{1}{{\sigma_{ij}}\,\sqrt{2\,\pi}} \int_{-\infty}^{{q}_i-{q}_j} e^{ \left (  \frac{-x^2}{2\,{\sigma_{ij}}^2} \right )} dx \, . \notag
\end{eqnarray}
%where $\Phi$ is the cumulative Normal distribution. 

% \begin{figure}[t]
% \centering
% \includegraphics[width=0.45\textwidth]{plots/difference_three_quality_scores}
% \caption{Probabilities of the distance between the conditions in Figure \ref{fig:three-quality-scores}. Differences are also normally distributed. }
% \label{fig:differences-three-quality-scores}
% \end{figure}

%The next step is to convert probabilities into distances in terms of quality scores. 
%Since scores are assumed to be normally distributed, 
 The mapping from probabilities into score differences is given by the inverse of $\Phi$ (know as the probit and shown in Figure~\ref{fig:cumulative-inv-norm}):
\begin{equation}\label{eq:mappingPQ}
 {q}_i - {q}_j = \sigma_{ij} \Phi^{-1}(P(r_i>r_j)).
\end{equation}
% The inverse cumulative distribution function $\Phi^{-1}$
% of the standard normal computes the number
% of standard deviations that $p_{ij}$ is from the mean.

{
Thurstone's model assumes that the noise parameter $\sigma$ is known and constant for all conditions, so that $\sigma_{ij} = \sigma \sqrt{2}$. However, we do not know its value. A common approach is to select $\sigma_{ij}$ so that a probability of $0.75$, in the midway between a random guess and being completely certain, is mapped to a score distance of 1 JOD unit. The difference of 2 JODs corresponds to the probability of 0.91 and so on. The inverse cumulative distribution crosses the value of 1 for $p_{ij}=0.75$ when the standard deviation $\sigma_{ij}$ is 1.4826.}

\subsection{Least-square distance solution}

Once that we have established the relation between probabilities and score differences, we can substitute $P(r_i>r_j)$ by the estimate $\hat{p}_{ij}$ in Eq. \eqref{eq:probfromC} {to obtain an estimate of the distance: 
\begin{equation}\label{eq:estmappingPQ}
 d_{ij} = \sigma_{ij} \Phi^{-1}(\hat{p}_{ij}).
\end{equation}
When these probabilities are transformed into score differences, we obtain the following distance matrix:
\begin{equation}
\mathbf{D}=\begin{bmatrix}
       0     & -2.7190  &    -Inf \\
    2.7190   &    0     & -1.0792 \\
       Inf   &  1.0792  &       0
\end{bmatrix}
\label{eq:distance-matrix}
\end{equation}}

Our aim is to find an estimation $\hat{q}$ such that the distances between the different scores closely resemble the distances in matrix $\mathbf{D}$. Such quality scores are often found by solving an optimisation problem of the form \citep{Engeldrum}:
\begin{equation}
\argmin_{\hat{q}_2,\ldots, \hat{q}_n} \sum_{i=1}^{n-1} \sum_{j=k}^n ((\hat{q}_i-\hat{q}_j) \, - \, d_{ij})^2.
\label{eq:mds-optim}
\end{equation}
This formulation is similar to the problem of multi-dimensional-scaling when we scale to a single dimension, except that our distances are signed. Since it is not possible to optimise the absolute score values given only distances between them, one of the scores is usually fixed (most commonly $\hat{q}_1=0$). %and the problem be solved for the remaining scores. 

Unfortunately, the solution of Eq. ~\eqref{eq:mds-optim} is unfeasible in our example because of the infinite values in $\mathbf{D}$. The two infinite values correspond to the cases when all observers gave the same (unanimous) answer and the probability is equal to 0 or 1. As the inverse cumulative Normal distribution reaches one of its asymptotes at 0 and 1, the corresponding distances in scores are infinite. The distance of plus or minus infinity is definitely an incorrect estimate, but it is also impossible to tell exactly what the true distance should be, given the data. Having unanimous answers is common in experiments, so it is highly important to devise a method to deal with those cases. 
Sometimes unanimous answers are ignored, but this removes valid observations from the data. In other cases the range of distances is restricted, for example to be between -3 and 3, but this introduces a bias in the estimate. {In the next section we present an optimisation method more suitable for these cases.}

\subsection{Maximum likelihood estimation}
\label{sec:mle}

A more elegant and robust solution for scaling is offered by Maximum Likelihood Estimation (MLE). Instead of minimising stress in distances in Eq. \eqref{eq:mds-optim}, MLE looks for the difference in quality scores that maximise the probability of observing our data $\mathbf{C}$.
%RFM: The text below is a repetition - commented out
%We can formally state our problem in terms of probability distributions. 
%We can again fix the quality score for the first condition at $0$ and assume that quality scores are random variables with Normal distribution $N$: 
%\begin{align}
%\hat{q}_1 & \equiv 0,
%\\
% r_k & \thicksim N(\hat{q}_k, {\sigma}) \quad \textrm{for} \quad k=1,\ldots,n \, , \end{align}
%where $\sigma =1.4826/\sqrt{2}$ (to ensure that the difference between two quality scores is equal to 1 JOD for 75\% preference). 
%Our goal will be to find the expected values $\hat{q}$ (which maximise the maximum likelihood). %For now, we can assume that they lie in the range 0 to $n-1$ and follow an uniform distribution $U$:
% \begin{equation}
% s_{prior} \thicksim U(0,n-1) \, .
% \end{equation}
% Note that the above assumption is our prior and not a constraint and therefore quality scores can achieve values higher than $n-1$ and lower than 0. 
To do so, we need to connect the quality differences with the data collected in the comparison matrix $\mathbf{C}$. If we know the true probability of selecting $O_i$ as better than $O_j$ ($P(r_i > r_j)$), the probability that $O_i$ was selected over $O_j$ in exactly $c_{ij}$ trials from the total number of $n_{ij}=n_{ji}=c_{ij}+c_{ji}$ trials is given by the binomial distribution:
\begin{eqnarray}\label{eq:binomial}
L(\hat{q}_i - \hat{q}_j|c_{ij},n_{ij}) &={{n_{ij}}\choose{c_{ij}}} \, P(r_i > r_j)^{c_{ij}} \,(1-P(r_i > r_j))^{n_{ij}-c_{ij}}  \\
&= {{n_{ij}}\choose{c_{ij}}} \, \Phi \left ( \frac{\hat{q}_i-\hat{q}_j}{\sigma_{ij}} \right )^{c_{ij}} \,\left (1-\Phi \left ( \frac{\hat{q}_i-\hat{q}_j}{\sigma_{ij}} \right ) \right )^{n_{ij}-c_{ij}} \notag
\end{eqnarray}
Note that, as shown earlier in Eq. \eqref{eq:prob-diff}, the probability $P(r_i > r_j)$ depends on the difference in quality scores and is given by the cumulative Normal distribution $\Phi$.
%The above formulation can be used to express the problem in a probabilistic programming language, such as BUGS or Stan, and used to find the posterior distribution via Markov chain Monte Carlo sampling. However, since our distributions are mostly normal, almost identical result can be found more efficiently by minimum likelihood estimation (MLE). 
%The MLE approach computes the likelihood (of observing data) given $\mathbf{D}$. 
%As said before, given $\hat{q}_i - \hat{q}_j$, we can compute $p_{ij}$ using the function shown in Figure~\ref{fig:cumulative-inv-norm}. 
%Given the probability $p_{ij}$ of observing condition $O_i$ as better than condition $O_j$ \eqref{eq:prob-diff}, the likelihood of observing the outcome of an experiment is given by the binomial distribution in \eqref{eq:biniomial-distr}. 

%where $C_{ij}$ is the number of times condition $i$ was selected as better than $j$ and $n_{ij}$ is the total number of comparisons of these two conditions ($n_{ij}=C_{ij}+C_{ji}$). 

%When we maximise this likelihood over computed pairs of distances in quality scores, the probability $P(r_i > r_j)$ explains the observed data (matrix $\mathbf{C}$) in the best possible manner.%and thus we can assign to this pair a distance in quality scores. %Note that we use upper-case ${L}$ to denote the likelihood of observing the collected data, and a lower-case $p$ to denote the probability of selecting one condition over another. 

To scale all compared conditions, we maximise the product of the likelihood for all pairs of conditions:
\begin{equation}
   \argmax_{\hat{q}_2,\ldots,\hat{q}_n} \prod_{i,j{\in}\Omega} L(\hat{q}_i - \hat{q}_j|c_{ij},n_{ij}) %= \argmax_{\hat{q}_2,\ldots,\hat{q}_n}\sum_{i,j} c_{ij} \cdot log(\Phi(\hat{q}_i - \hat{q}_j))
\label{eq:mle}
\end{equation}
%where $L_{ij}$ is given in \eqref{eq:binomial}. 
where $\Omega$ is the set of all pairs for which at least one comparison has been made: $c_{ij}+c_{ji}>0$. 
%The problem can be efficiently solved using numerical methods, since there is no closed-form solution. 
Note that, in practice, it is more convenient to maximise the log of the likelihood function.

\begin{figure}[t]
\centering
\includegraphics[width=0.6\textwidth]{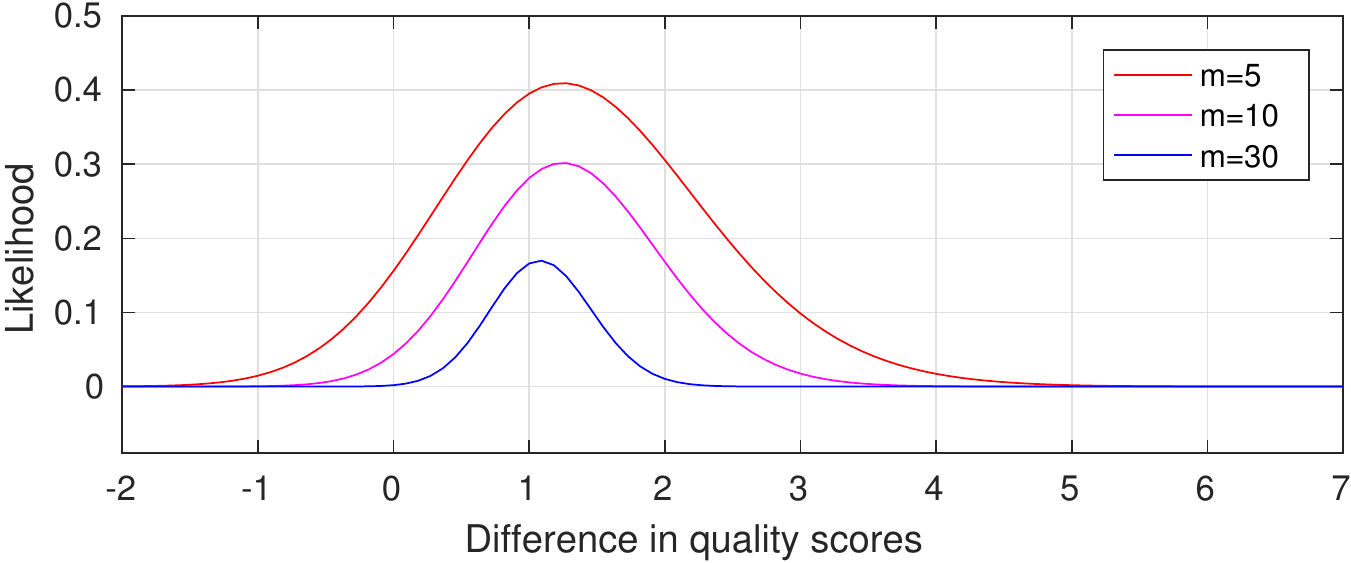}
\caption{The likelihood function of JOD distance for three sample sizes and when $p_{ij}=0.75$. Note that the range of likely distance values gets smaller with the number of samples.}
\label{fig:likelihood-certainty}
\end{figure}

\begin{figure}[t]
\centering
\includegraphics[width=0.6\textwidth]{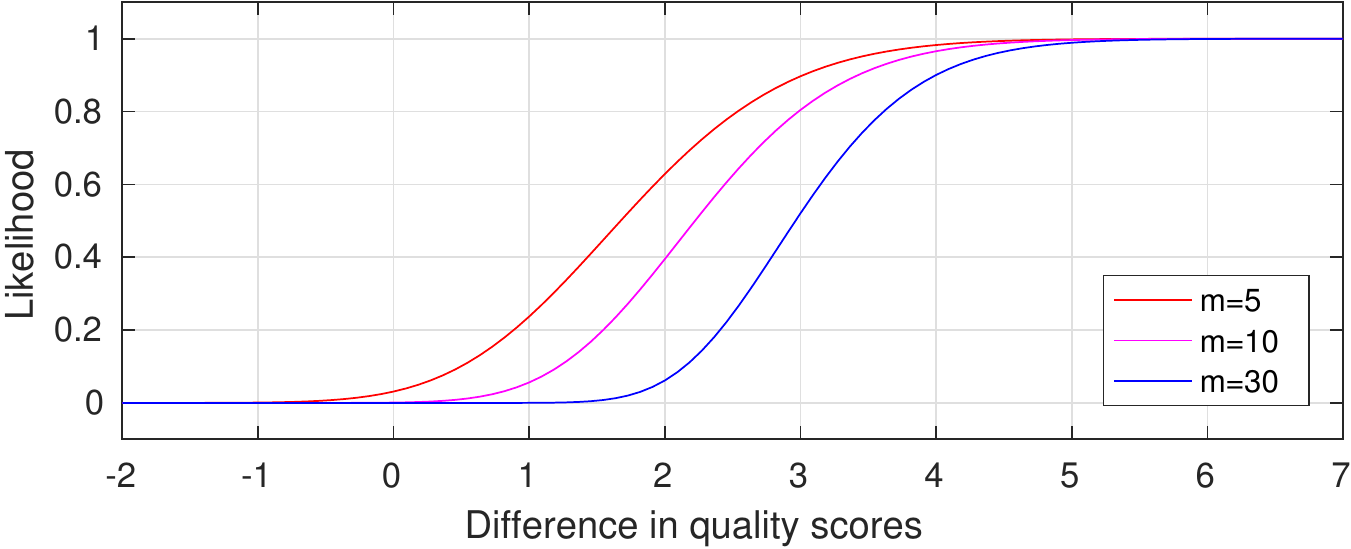}
\caption{The likelihood function of JOD distance for three sample sizes and when one condition is always selected as better ($p_{ij}=1$).}
\label{fig:likelihood-unanimous}
\end{figure}

Solving MLE in Eq. \eqref{eq:mle} has a number of advantages over the least square distance solution in Eq.\eqref{eq:mds-optim}:
\begin{itemize}
\item MLE accounts for the number of comparisons and thus the measure of confidence we have in our data. Figure~\ref{fig:likelihood-certainty} shows the likelihood for three sample sizes for $p_{ij}=0.75$. The larger the sample, the narrower is the range of likely differences between the scores. This property of the MLE solution is in particular useful when the experimental design is not balanced. %and some conditions are compared more times than the others. 

\item MLE solution (almost) gracefully handles the cases with unanimous answers. Figure~\ref{fig:likelihood-unanimous} plots the likelihood $L$ as the function of difference in quality scores, for the case when $p_{ij}$ is equal to 1 and the number of observers $m$ is 5, 10 and 30. In each case, the most likely distance is greater than 5, but there is also a likelihood of a smaller distance, especially when $m$ is small. 

\item MLE allows us to work with incomplete experimental designs, when only a subset of pairs is compared.%(e.g. using a sorting algorithm to reduce the number of required comparisons \citep{Silverstein2001}).

\end{itemize}

\section{Statistical analysis}
\label{sec:stat-analysis}

Since any experiment gives only estimates of the true quality values, it is important to analyse and report the level of uncertainty in the data. In this section we show how to compute confidence intervals and test for statistical differences. 

 %If data are not proved to lie in a given interval with 95\% probability, then we might suspect that observed differences are due to a measurement error, and not to a genuine effect.

\subsection{Confidence intervals}
\label{sec:confidence-intervals}

Computing confidence intervals for scaled quality scores using analytical methods is difficult because multiple conditions influence each other. {The original formulation of Thurstone Case V does not allow the computation of confidence intervals. Different authors have change the base model to account for this \citep{Montag2003}, but this is at the cost of the simplicity of the model.} However, confidence intervals can be computed using numerical methods, e.g. resampling (see for example ch. 18.1 in \citep{Howell2009}).
Resampling is generally used as a statistical method for estimating the sampling distribution. It represents a robust alternative to inference based on parametric assumptions
when those assumptions are in doubt. A common example is the use of the bootstrapping technique. This method always resamples from the sample, therefore relying on the generation of pseudo-samples from the sample collected. Given a measured sample (result of a pairwise comparison experiment), we generate a new sample of the same size by randomly replicating data for some participants and removing data for others.  The procedure is know as random sampling with replacement. 
 To compute confidence intervals, a large number of pseudo-samples in generated (usually more than 500), then each sample is scaled using the MLE method from Section~\ref{sec:mle}, and finally the  2.5-th and 97.5-th percentiles of JOD values are computed for each condition across all samples. This gives the 95\% confidence intervals for the mean JOD scores.

\begin{figure}[t]
\centering
\includegraphics[width=0.7\textwidth]{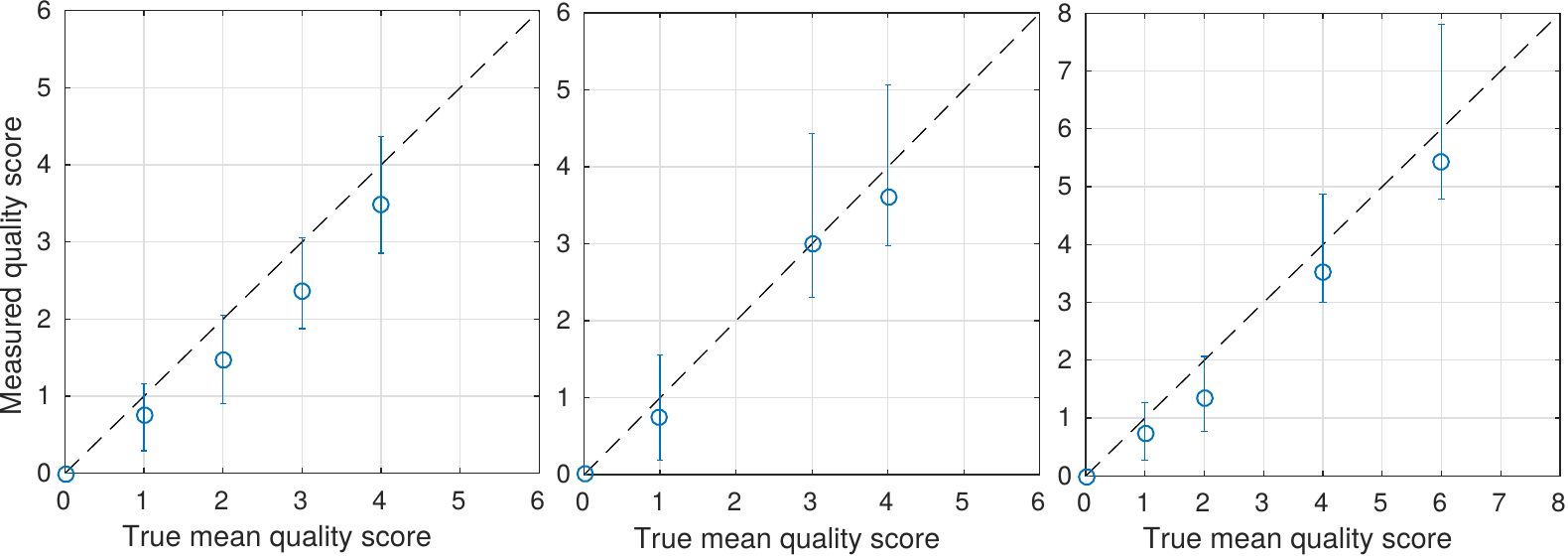}
\caption{Results of the simulated experiments, in which the pairwise comparison results have been generated by randomising scores using probability distributions similar to those shown in Figure~\ref{fig:three-quality-scores}. The known means of those distributions are shown on the x-axis and the results of JOD-scaling on the y-axis. The error bars denote 95\% confidence intervals computed by bootstrapping.}
\label{fig:sim-exps}
\end{figure}

Figure~\ref{fig:sim-exps} shows three examples of confidence intervals computed for simulated experiments. 
Assuming a set of fixed true scores, we can simulate the randomness of observers' judgments by drawing simulated answers from distributions, such as those shown in Figure~\ref{fig:three-quality-scores}. In our examples, ten virtual observers (n=10) performed three repetitions (t=3) of the experiment, in which all pairs were compared.
There are a few conclusions that can be drawn from the plot: 
\begin{itemize}
\item Confidence intervals are larger for quality scores that are farther from the reference point 0. Since the absolute scores are estimated from distances between the pairs, the estimation error between the first and second condition is propagated to the third condition, and so on.

\item Confidence intervals become larger as the distance between conditions increases. Intuitively, Figure~\ref{fig:cumulative-inv-norm} shows that larger distances are projected onto smaller differences in probability. Thus, when a JOD distance is large, a small error in the estimation of probabilities can cause a large error in estimated distance. 
\end{itemize}

\begin{figure}[t]
\centering
\includegraphics[width=0.5\textwidth]{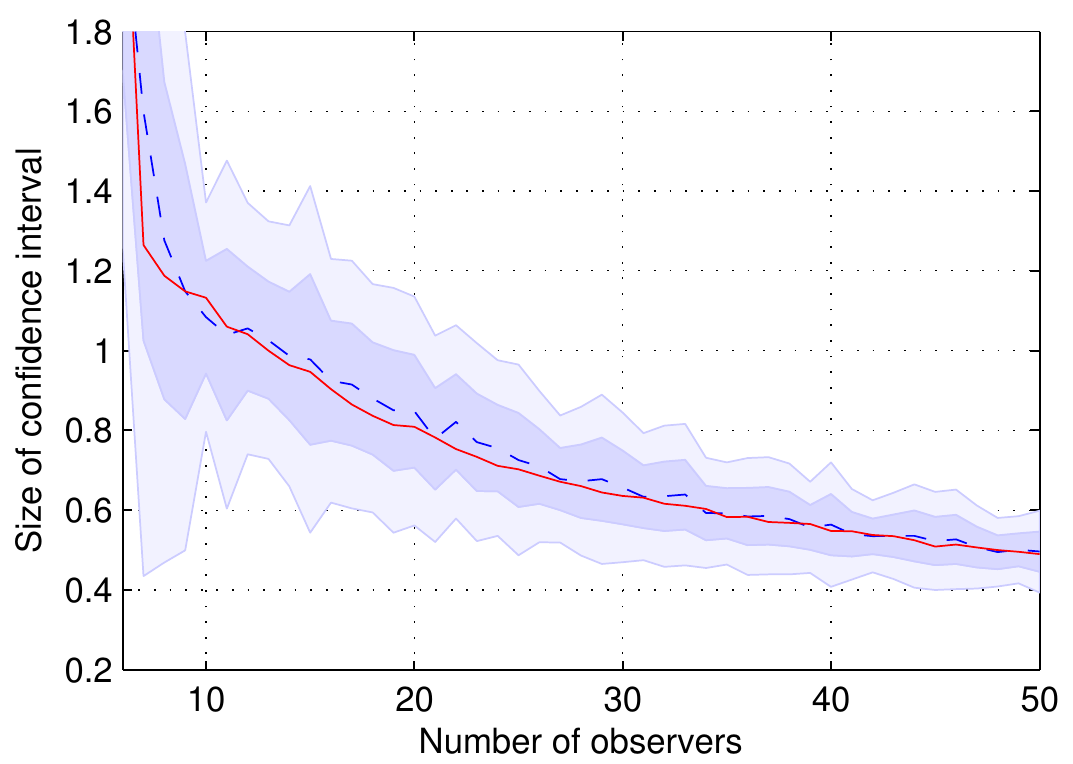}
\caption{The confidence interval computed from experiment data via bootstrapping compared with the actual confidence interval found in a Monte-Carlo simulation (10,000 simulated experiments). The bootstrapping on average (dashed-blue line) follows very closely the true confidence interval (red continous line). The darker and brighter blue-shaded area represent the standard deviation and 95\% interval for 50 experiment runs for which bootstrapping was computed. }
\label{fig:bootstrap-error}
\end{figure}

To analyse how accurate bootstrapping is for estimating confidence intervals in our problem we analyse its performance in a simulation, where true confidence intervals can be estimated with high precision. We assume we know that the true quality scores are $q=(0, 1, 2, 3, 4)$. Then, we simulate 10,000 runs of an experiment by randomising answers of a certain number of observers (adding random Gaussian noise $N(0,1.4826)$ to $q$), generating corresponding comparison matrices and running our scaling method. We compute the mean size of the confidence interval (mean of the distance between 97.5th percentile and the mean; and the distance between the mean and 2.5th percentile) and plot it for experiments with different number of observers in Figure~\ref{fig:bootstrap-error} (red continuous line). Then, we use the same procedure to simulate 50 experiments (for each number of observers) for which we run bootstrapping and compute the mean size of the confidence interval in the same way. The distribution of bootstrapping results is shown as the blue-shaded areas and the blue dashed line in Figure~\ref{fig:bootstrap-error}. It can be seen that on average bootstrapping gives us a correct estimate. However, we need to keep in mind that bootstrapping is just an estimate, and the computed interval can be easily both under- and over-estimated, especially when the number of observers is small. Therefore, we need to have limited confidence even in the confidence intervals.  

\subsection{Statistical difference between two conditions}

%\textcolor{red}{Explain how we test whether differences are statistically significant}

The analysis of confidence intervals for pairwise comparison data is more complicated than for a typical direct rating experiment because the computed JOD values are not independent. Since all conditions are ``linked'' to each other by pairwise comparisons, changing the value of one condition will ``push'' the values of all directly or indirectly linked conditions. This correlation between conditions can be captured in a covariance matrix $\Sigma$, such as one shown below:
\begin{equation}
\Sigma=\begin{bmatrix}
          0 &        0  &       0 \\
         0  &  0.431   &  0.486 \\
         0  &  0.486   &  0.683
\end{bmatrix}
\label{eq:sigma-matrix}
\end{equation}
The first row and column have 0s because $O_1$ is always fixed at 0 and cannot vary. Values $\Sigma_{22}=0.431$ and $\Sigma_{33}=0.683$ represent variance for $O_2$ and $O_3$. The value $\Sigma_{23}=\Sigma_{32}=0.486$ represents the variance between a pair of conditions. If we want to reject $H_0$ that the difference in JOD scores between two conditions is 0, we need to compute the variance for that difference as:
\begin{equation}
  v_{ij}=\Sigma_{ii}+\Sigma_{jj} - 2\,\Sigma_{ij}\ .
\end{equation}
Using the variance and the difference in JOD scores, a two-tailed test can be used to test $H_0$ \citep{David1963} for a given level of confidence.

% 
% \begin{figure}[t]
% \centering
% \includegraphics[width=0.35\textwidth]{figures/triangles.pdf}
% \caption{Graphical representation of the scaling solution. Red points represent conditions, and they are only connected to their neighbours, as these are usually the comparisons in which we are most interested. Blue solid lines represent statistically significant differences, as opposed to red dashed lines. The x-axis shows the scaling solution. }
% \label{fig:triangles}
% \end{figure}

\section{Finite distance prior}
\label{sec:prior}

% \textcolor{red}{What is the theory behind the prior? Can we link the number of observers, and thus $n_{ij}$ using the binomial, to the probability of having unanimous answers, showing perhaps that unanimous answers are of course much more probable with less observers? }

Unanimous answers are problematic for scaling methods as they put no upper bound on the distance between two conditions and thus introduce a bias in the estimation. 
The problem is most noticeable when the sample size (number of observers) is small. This is because i) the probability of having unanimous answers increases with few observers, and ii) the
smaller the sample, the wider is the range of likely differences
between the scores (see Figure \ref{fig:likelihood-certainty}). However, the scaling can be made more robust by adding a simple distance prior to the likelihood function. %which we present in this section. 

 \begin{figure*}[t]
\centering
\includegraphics[width=0.32\textwidth]{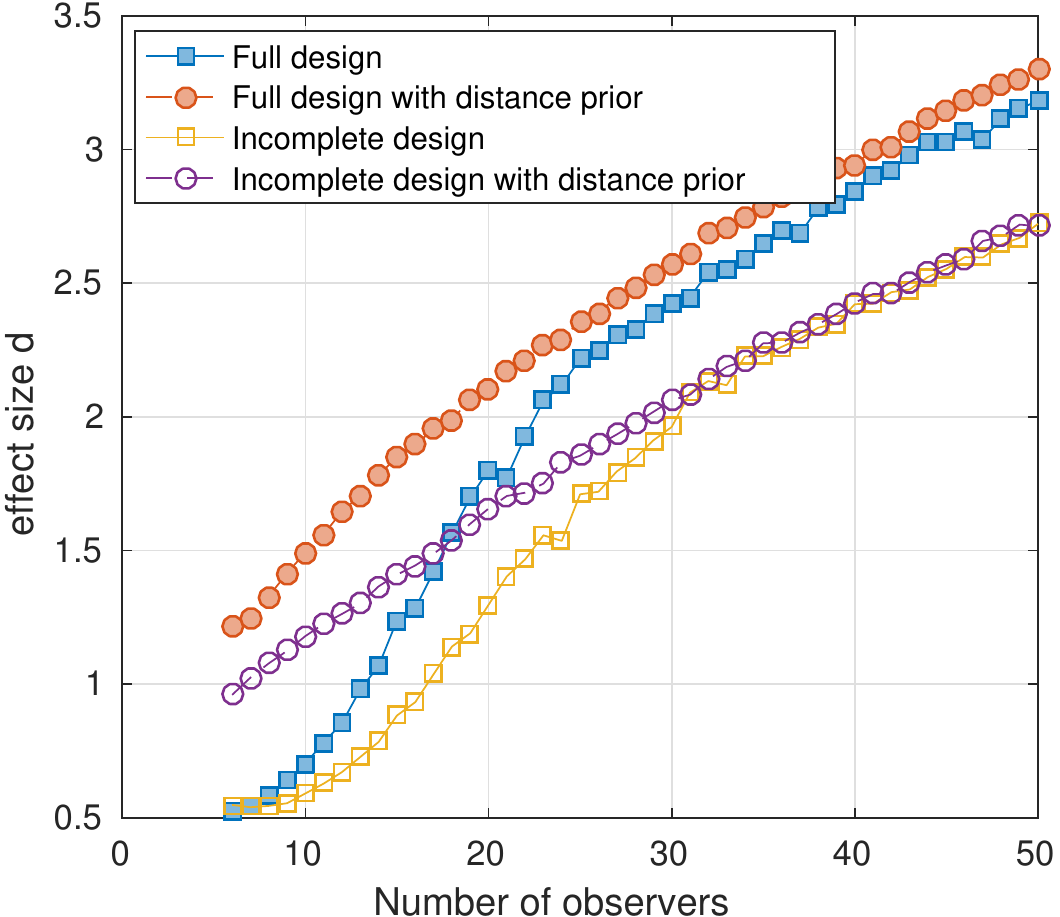}
\includegraphics[width=0.32\textwidth]{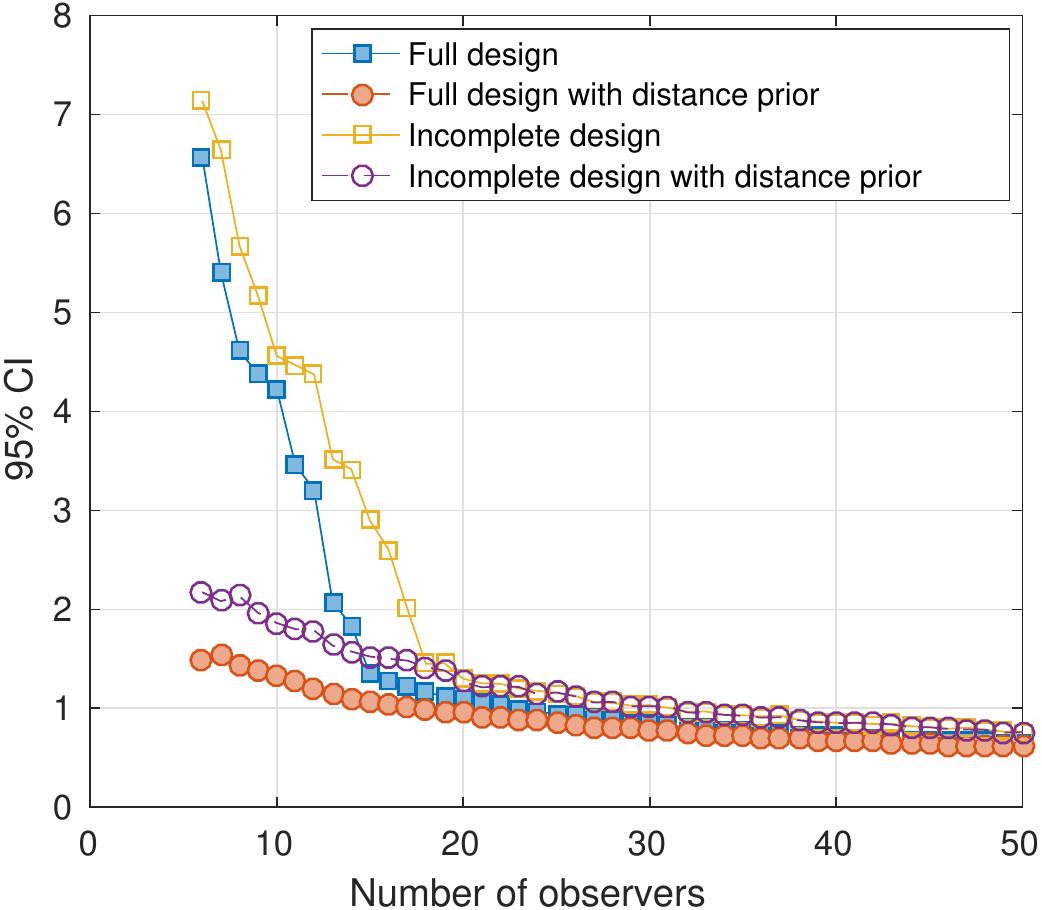}
\includegraphics[width=0.33\textwidth]{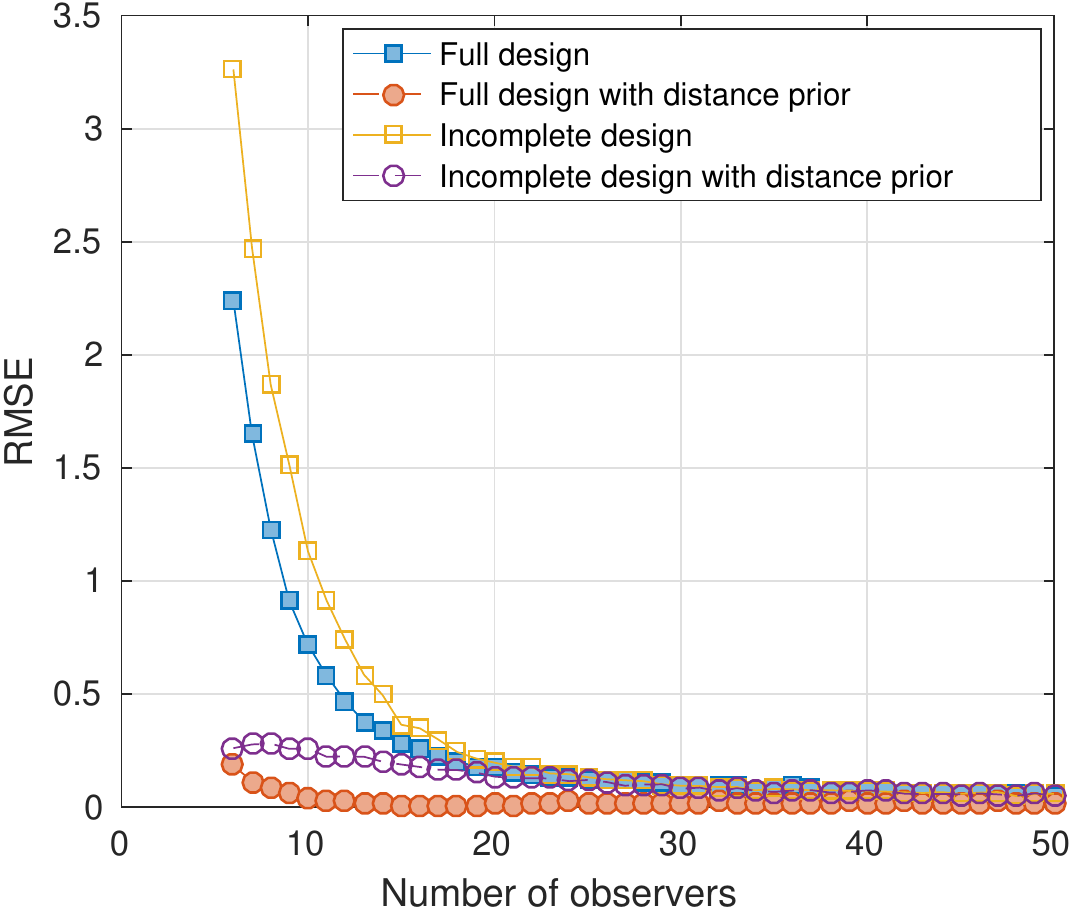}
\caption{The difference in precision with and without the distance prior. Each plot represents a different measure (see text for details). }
\label{fig:simulated-with-prior}
\end{figure*}

% \begin{figure*}[t]
% \centering
% \includegraphics[width=0.31\textwidth]{simulation/simulated_with_prior_dNew}
% \includegraphics[width=0.31\textwidth]{simulation/simulated_with_prior_ciNew}
% %\includegraphics[width=0.22\textwidth]{simulation/simulated_with_prior_bias}
% \includegraphics[width=0.32\textwidth]{simulation/simulated_with_prior_rmseNew}
% \caption{The difference in precision with and without prior. Each plot represents a different measure (see text for details). }
% \label{fig:simulated-with-prior}
% \end{figure*}

\begin{figure}[t]
\centering
\includegraphics[width=0.7\textwidth]{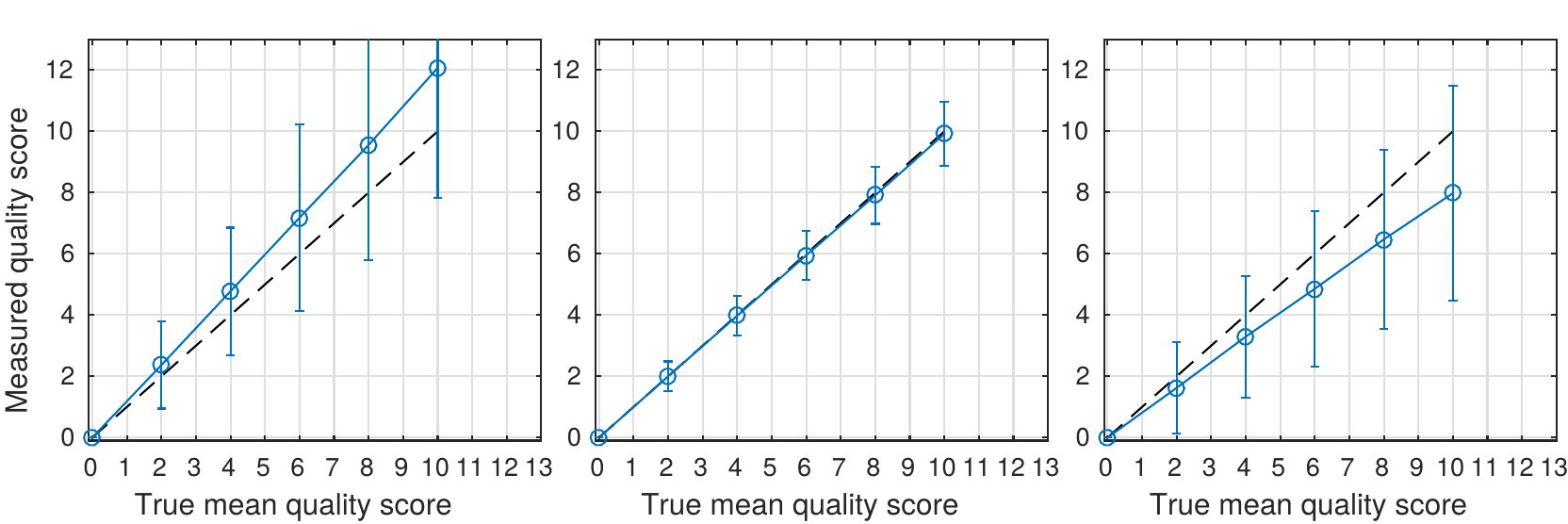}
\caption{Averaged JOD values from running 1000 simulated (independent) experiments. Left: when the original MLE formulation from Eq. \eqref{eq:mle} is used, the mean shows bias towards higher JOD differences. Middle: when the prior is added in Eq. \eqref{eq:mle-prior}, bias is reduced. Right: when all comparison with unanimous answers ($p_{ij}=0$ or $p_{ij}=1$) are removed, the bias is tipped to the other side, i.e. JOD distances are under-estimated. }
\label{fig:sim-exp-bias}
\end{figure}

\begin{figure}[t]
\centering
\includegraphics[width=0.6\textwidth]{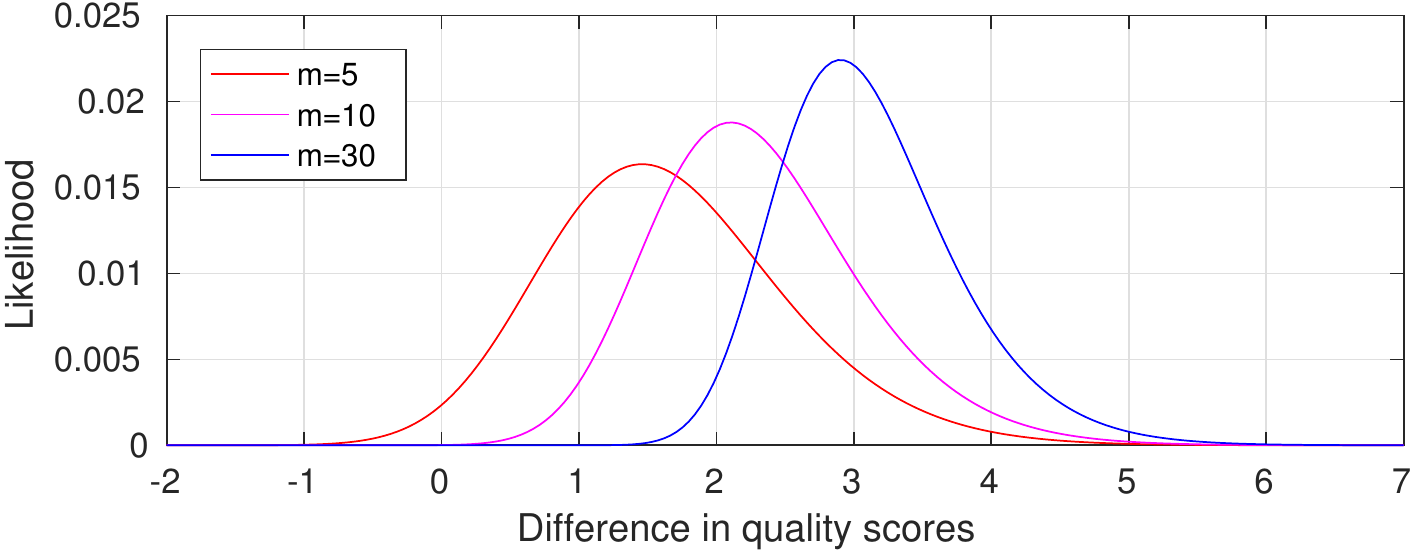}
\caption{The likelihood function modulated by a distance prior (posterior) for the case when all answers are unanimous.} %The prior compensates for bias in estimates.}
\label{fig:likelihood-unanimous-prior}
\end{figure}

 \begin{figure*}[t]
\centering
\includegraphics[width=0.75\textwidth]{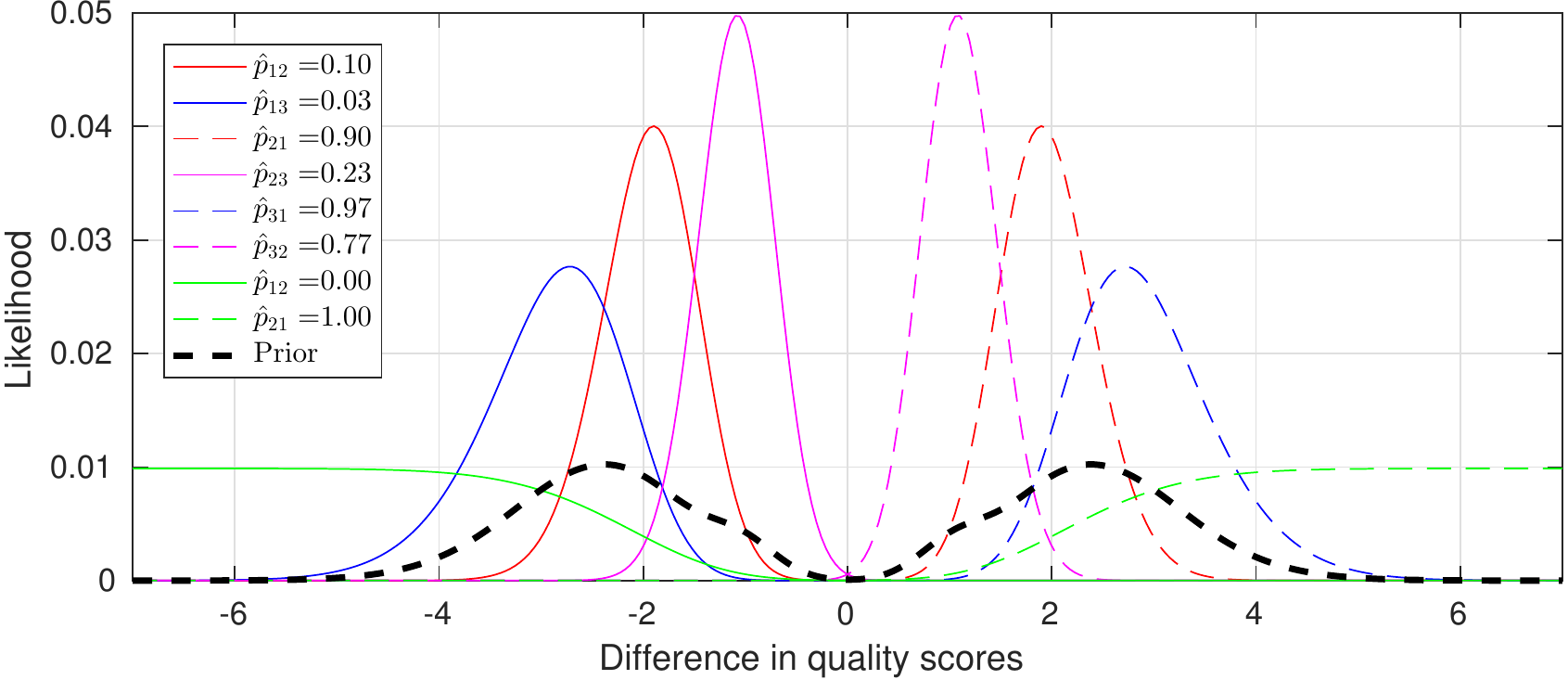}
\caption{The proposed prior (dashed-line) as the normalised sum of probabilities of observing a difference for all compared pairs of conditions. The distribution are computed for our toy-example from Eq. \eqref{eq:measurement-matrix}. }
\label{fig:differences-three-quality-scores}
\end{figure*}

This problem is easier to see on the example shown in Figure~\ref{fig:sim-exp-bias}, where 1000 runs of an experiment are simulated for the true scores of $q = (0, 2, 4, 6, 8,10)$. If there was no bias in the method, the experiments should, on average, give the correct answer, exactly on the black-dashed line in the left side of Figure~\ref{fig:sim-exp-bias}. However, due to the bias, the measured scores are larger than the true scores. The reason for that are the cases of unanimous answers, which put no upper bound on the distances between conditions. The likelihood for those cases, as shown in Figure~\ref{fig:likelihood-unanimous}, ensures that distances are larger than a certain value, but they do not restrict the maximum distance values. Such cases are pushing conditions on the quality scale away from each other. It may seem that it would be much easier to ignore the cases of unanimous answers from the comparison matrix. However, as we show in the right plot in Figure~\ref{fig:sim-exp-bias}, this leads to under-estimated JOD values.

% \begin{figure}[t]
% \centering
% \includegraphics[width=0.35\textwidth]{plots/p_dist_uniform}
% \includegraphics[width=0.35\textwidth]{plots/p_dist_triangle}
% \caption{The difference of two random variables drawn from an uniform distribution (shown at the top) results in the triangle distribution shown at the bottom.}
% \label{fig:uniform-triangle-dist}
% \end{figure}
\begin{figure*}[t]
\centering
\includegraphics[width=0.32\textwidth]{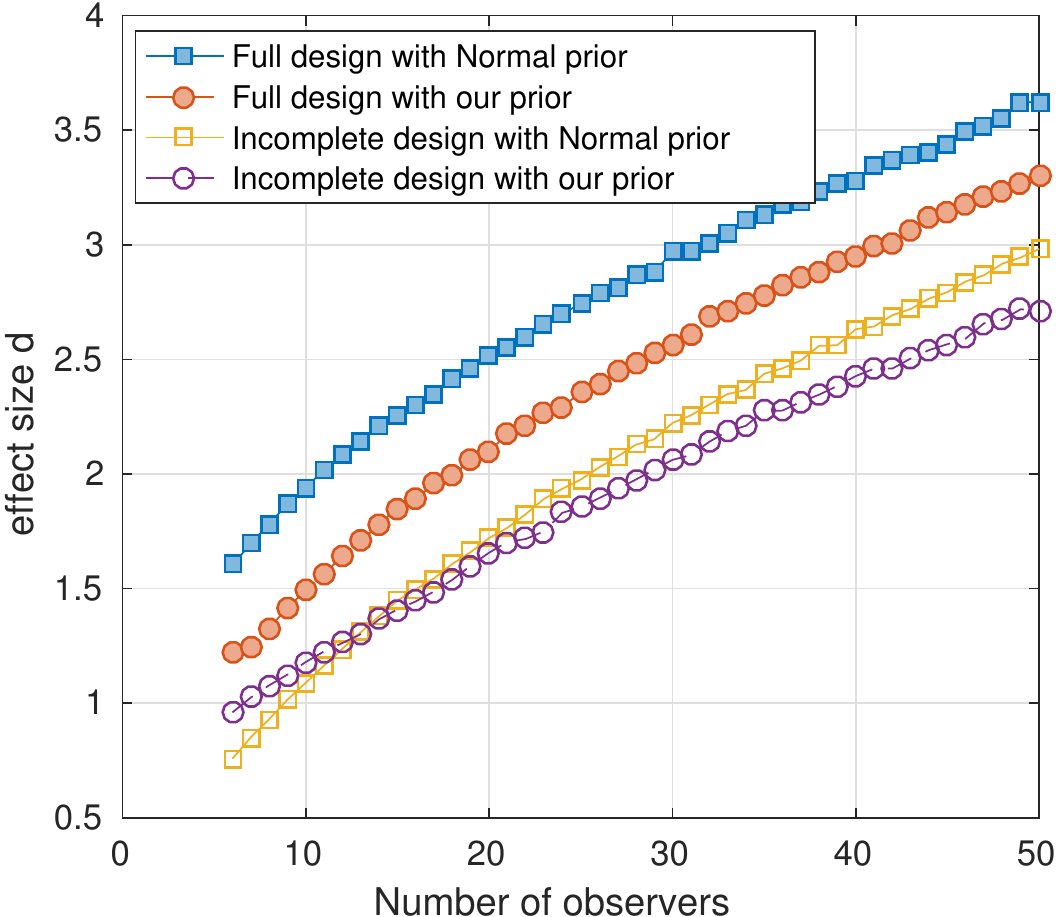}
\includegraphics[width=0.32\textwidth]{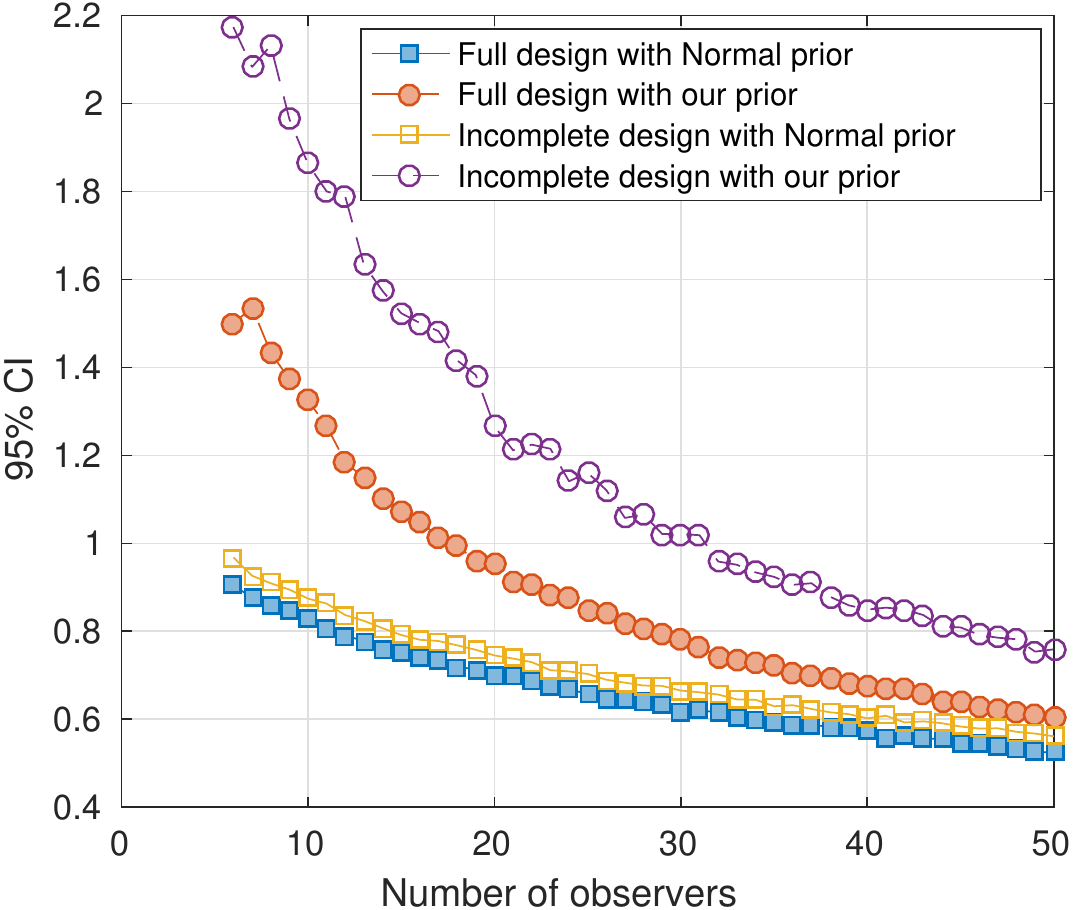}
\includegraphics[width=0.33\textwidth]{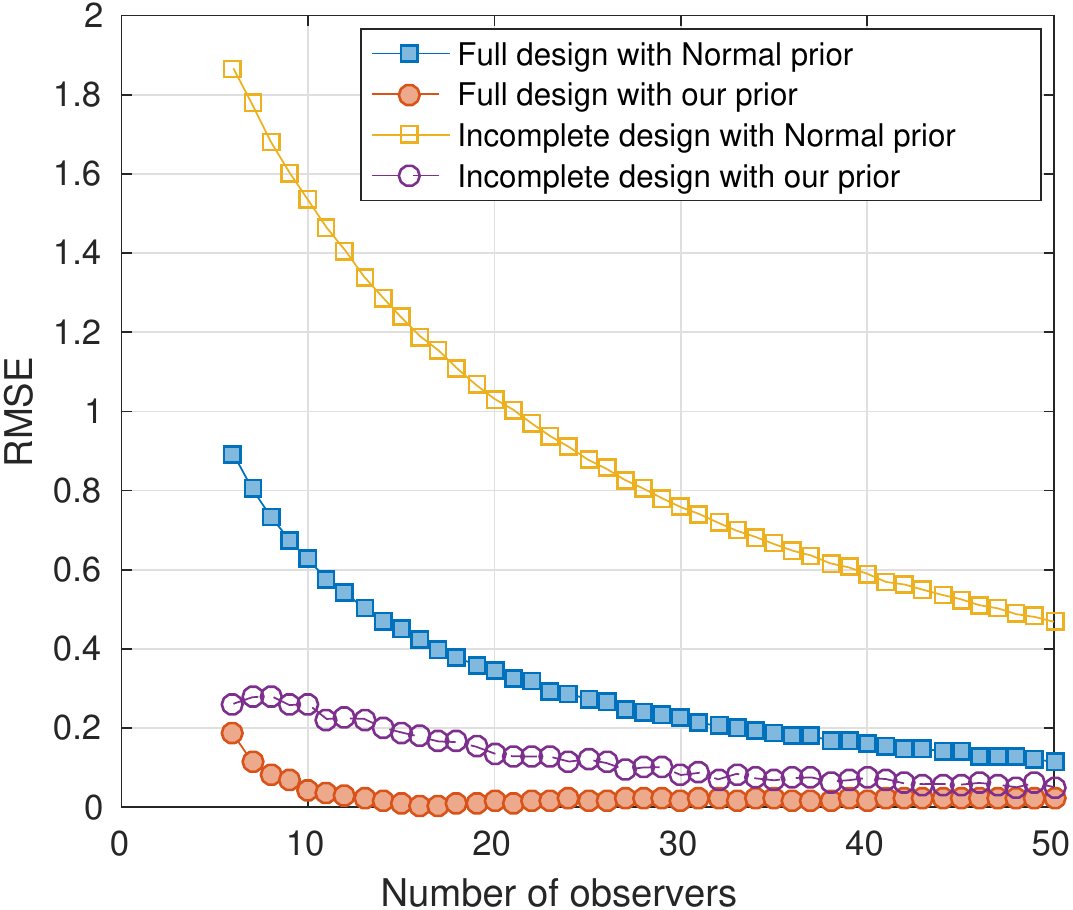}
\caption{The difference in precision with the prior in \citep{Tsukida2011} for the previously presented metrics. }
\label{fig:simulated-tutorial}
\end{figure*}

% \begin{figure*}[t]
% \centering
% \includegraphics[width=0.21\textwidth]{simulation/simulated_two_priors_d}
% \includegraphics[width=0.21\textwidth]{simulation/simulated_two_priors_ci}
% \includegraphics[width=0.22\textwidth]{simulation/simulated_two_priors_bias}
% \includegraphics[width=0.22\textwidth]{simulation/simulated_two_priors_rmse}
% \caption{The difference in precision with the previously proposed prior (based on the uniform distribution). }
% \label{fig:gaussian-vs-uniform}
% \end{figure*}

%Such bias can be reduced by introducing prior knowledge into our optimisation process. 

Although the likelihood functions in Figure~\ref{fig:likelihood-unanimous} allow distances between conditions to be infinity, we know that in practice all distances are finite and usually moderate numbers. Such knowledge of finite distances will be our prior. We can define as our prior the likelihood of observing a particular distance in quality scores for any randomly selected pair of conditions. Such likelihood for a given pair of conditions is expressed in Eq. ~\eqref{eq:binomial}. Given our toy-example comparison matrix from Eq. ~\eqref{eq:measurement-matrix}, we plot the likelihood for all pairs of conditions in Figure~\ref{fig:likelihood-unanimous-prior}. The probability of observing any difference is a normalised sum of all plotted probabilities. The problem is, however, that the likelihood for unanimous answers ($\hat{p}_{12}$ and $\hat{p}_{21}$, green lines in the plot) has infinite support and thus cannot be normalised. To avoid this issue, we transform this answer to the closest non-unanimous response. After this step, we can compute the probability of observing a distance $z$ between any two random conditions as:
\begin{equation}
l(z) = \frac{1}{|\Omega|}\sum_{i, j \in \Omega} L(z| c^*_{ij},n_{ij}),
\end{equation}
where $c^*_{ij}$ is the non-anonimous closest version of $c_{ij}$. The main term in the sum is given by Eq. \eqref{eq:binomial} and our prior depends on the estimated distances in the current iteration of the optimisation method and changes in an iterative manner. This probability is shown as a dashed-black line in Figure~\ref{fig:likelihood-unanimous-prior}. It shows that the most probable difference between two randomly chosen conditions is about 2.5 JODS, and the support of this probability function is finite. We can add our distance prior to the likelihood function from Eq. \eqref{eq:mle}:
\begin{equation}
 \argmax_{\hat{q}_2,\ldots,\hat{q}_n} \prod_{i,j{\in}\Omega} L(\hat{q}_i - \hat{q}_j|c_{ij},n_{ij}){\cdot}(l(\hat{q}_i - \hat{q}_j)+\gamma) \ .
\label{eq:mle-prior}
\end{equation}
Note that this is just a prior modulating distances, not a constraint. To allow the selection of other distances, we add a small offset of $\gamma = 0.1$ to our prior. 
The centre plot in Figure~\ref{fig:sim-exp-bias} demonstrates how the bias is reduced when the prior is included in the likelihood function. %Note that this formulation considers only compared conditions and takes into account the number of comparisons and the confidence we put in our data.

Figure~\ref{fig:likelihood-unanimous-prior} shows the likelihood function from Figure~\ref{fig:likelihood-unanimous} when it is multiplied by the prior. The likelihood has no longer plateau and has a single maximum, which also improves the stability of the optimisation. 
%It can be seen that the distance between the two conditions increase with the number of observers and the range of possible distances decrease, as we have more confidence.

To evaluate the improvement in estimates brought by the prior, we analyse how the precision of the estimation varies with the number of observers. We perform a Monte Carlo simulation of $10,000$ runs for the true quality scores $q=(0, 1, 2, 3, 4)$ and with the same assumption as for estimation of confidence intervals in the Section~\ref{sec:confidence-intervals}. We run the simulation for both complete design (in which we compare all conditions) and incomplete design (in which only nearest neighbours are compared). For each simulation we obtain a set of $10,000$ estimated quality scores $\hat{q}$, which we aim to compare to the true quality scores in $q$. We define the mean for our estimation of $q_i$ as $\bar{q}_{i}$.
The results are shown  in Figure~\ref{fig:simulated-with-prior} for three different measures:
\begin{itemize}
\item Effect size $d$: ratio of the difference between estimated quality scores divided by the standard deviation of the estimation error:
\begin{equation}
d = \frac{1}{n-1}\sum_{i=1}^{n-1} \frac{(\bar{q}_{{i+1}} - \bar{q}_{i})}{\sigma_{\hat{q}_i}} \, ,
\end{equation}
where $\sigma_{\hat{q}_i}$ is the standard deviation of each individual estimated result from the mean of the distribution $\bar{q}$.
%We report effect size as the mean $d$ value for all pairs of neighbouring conditions. 
The effect size \citep{David1963} is a useful measure of the sensitivity of an experimental method, computing whether it can detect a difference between a pair of conditions and prove its statistical significance. 
\item The average size of 95\% confidence interval, computed as in Section~\ref{sec:confidence-intervals}.
%\item Bias, which represents the average difference between the true and estimated quality scores: $b = \frac{1}{N} \sum_{i=1}^N (\bar{q}_i - q_i )$. If a scaling method shows a positive bias, it means that it is more likely to overestimate JOD distances than to underestimate them. %If a scaling method has no bias,  we should be equally likely to over-estimate as under-estimate distances, thus resulting in the bias equal to 0. %\textcolor{red}{Is it enough with these measures? Don't we need something similar to MSE?}
\item {Root Mean Squared Error} (RMSE), which measures the deviation from the ground truth, as follows: RMSE$= \sqrt{ \frac{1}{n-1}\sum_{i=2}^{n} (q_i - \hat{q}_i)^2}$. 
\end{itemize}
Figure~\ref{fig:simulated-with-prior} shows how the measures improve as we increase the number of observers. It also shows that both the RMSE and the confidence intervals can be very large if the number of observers is less than 20. The proposed distance prior significantly improves accuracy and robustness of estimation, specially for small samples.

To demonstrate the challenge of selecting the right prior, we compare scaling using our distance prior to the prior proposed in \citep{Tsukida2011}. The authors introduced a prior that assumed the quality scores to be drawn from a normal distribution. Figure~\ref{fig:simulated-tutorial} shows that even though their prior strongly reduces confidence intervals (as most priors do), it also introduces a large error in the estimates (large RMSE).

%  $r_i \sim N(q_i, \sigma)$, and so distances between conditions $O_i$ and $O_j$ are also represented by normal random variables $r_{ij} = (r_i - r_j) \sim N(\mu_{ij}, \sigma_{ij})$, where $\mu_{ij} = q_i - q_j$ and $i < j$, ${\forall i,j\in \Omega}$. $r_i$ was assumed to be independent of $r_j$ for all $i \neq j$. However, now, the new variables $r_{ij}$ are not independent. The mean of all these distances between conditions give us a new random variable $z$ that represents the probability of any distance between two conditions, where 
% \begin{equation}
%  z \sim N \left (\frac{1}{{{n}\choose{2}}} \sum_{i<j} \mu_{ij}, \sqrt{ \sum_{i<j} \sum_{k<l} \frac{1}{{{n}\choose{2}}}^2 Cov(r_{ij},r_{kl}})\right),
% \end{equation}
% where all $i$, $j$, $k$ and $l$ go from 1 to $n$. $Cov(r_{ij},r_{kl}})$
% \begin{eqnarray}
% \sigma_z = \sum_{i=\{1, \ldots, n\},j>i} \left (\frac{1}{{{n}\choose{2}}} \right )^2 \cdot \sigma^2_{ij} + 2 \; \sum_{h=\{1, \ldots, n\},f>h} \\ \left (\frac{1}{{{n}\choose{2}}} \right )^2 Cov(r_i - r_j, r_h - r_f),
% \end{eqnarray}

\section{Outlier detection}
\label{sec:outlier}

In practice, some observers may not fully understand or follow the instructions of the experiment, in particular in less controlled crowdsourcing experiments. It is important to detect these observers that fall outside of the overall pattern because their answers can push the scaling towards an incorrect solution. This section presents a new method to detect those outlier observers. Note that this approach is only intended to support the experimenter, who makes the final decision on whether the observer should be considered an outlier and removed from the dataset.
%In this case, however, we have no true ground data, just information collected from different observers, which we assume that captures the true trend of the underlying quality scores. 

To indicate if a specific observer can be considered as an outlier, we compare her/his answers to the rest of the sample. First, we exclude a given observer (one by one) from the dataset and use MLE method to find the scaled distances and thus probabilities $P(r_i > r_j)$. Given these probabilities for the rest of the sample, we use the product of likelihoods (Eq. ~\eqref{eq:binomial}) to calculate the probability of observing the answers of the considered observer. If the considered observer is consistent with the rest of the population, the corresponding probability will be high. In practice, we use the sum of logarithmic likelihoods as it not only simplifies the subsequent analysis, but it also helps numerically, since the product of a large number of small probabilities can easily underflow the numerical precision of floating point numbers.

Different rules can be used to detect outliers, most of them taking into account the distance to a central measure of the distribution and the range of the data. In our case we consider J. Tukey's rules on quartiles. We transform log-likelihoods into the scores, which express the distance to the central range of the distribution in the multiples of the interquartile range. The interquartile range is the distance between $75^{th}$ and $25^{th}$ percentiles. We only consider outliers on the left side of the distribution, i.e. cases which show a significantly low likelihood of belonging to the sample, computing the distance to the first quartile. %Figure~\ref{fig:outlierDetectionPlot} represents an example showing the scores obtained by the presented procedure for all observers for the \citep{video TMO evaluation} dataset used in Section~\ref{sec:example-scaling}. The black dashed line represents the customary threshold used to detect outliers, meaning that there is an observer that could potentially be an outlier. 

%We only consider outliers on the left side of the distribution (i.e. cases which show a significantly low likelihood of belonging to the sample). This is, a score $A_u$ is considered to be an outlier if it lies on the left side of the distribution of the scores $A$ and it is more than $1.5\cdot IR$ times away of the first quartile $Q_1$, where $IR$ represents the interquartile range. 
 
% \begin{figure}[t]
% \centering
% \includegraphics[width=0.35\textwidth]{plots/likelihood_plot_outlier}
% \caption{Log-likelihood scores used to detect outliers obtained for the observers in the \emph{video TMO evaluation} dataset. }
% \label{fig:outlierDetectionPlot}
% \end{figure}

We ran a series of Monte-Carlo simulations to determine how the presence of an outlier affects the results of scaling and whether our method could be used to automatically determine outliers. As expected, an outlier can introduce the highest error when the number of valid observers (non-outliers) is small. But more interestingly, we observed that the outliers are more difficult to detect when the number of repetitions $t$ is small. Thus, we recommend that each observer repeats the same comparisons at least 3 times. When investigating the actual (non-simulated) datasets from previous papers, {we found that given the subjectivity of the experiments this automatic criteria for detecting outliers might not be always accurate}. Therefore, we recommend leaving this decision to an experimenter, who should investigate answers of flagged observers whose outlier scores are high (as discussed in Section~\ref{sec:example-scaling}).

\section{Practical issues}
\label{sec:practical}

In this section we explore three relevant issues concerning pairwise comparison experiments: the comparison between complete and incomplete designs, the distance between quality scores and the allowance of ties in the experiment.

\subsection{Complete or incomplete design}
\label{sec:fullvsincomplete}

\begin{figure*}[t]
\centering
\includegraphics[width=0.32\textwidth]{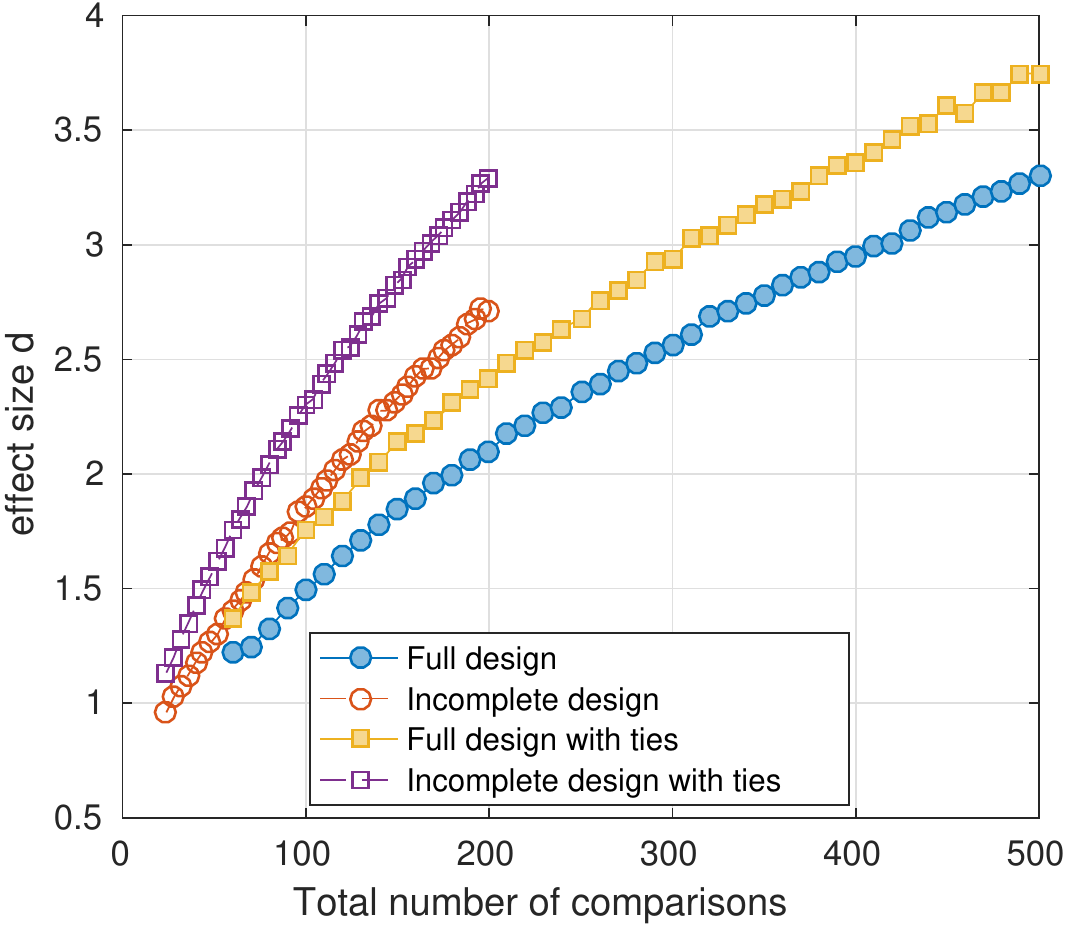}
\includegraphics[width=0.32\textwidth]{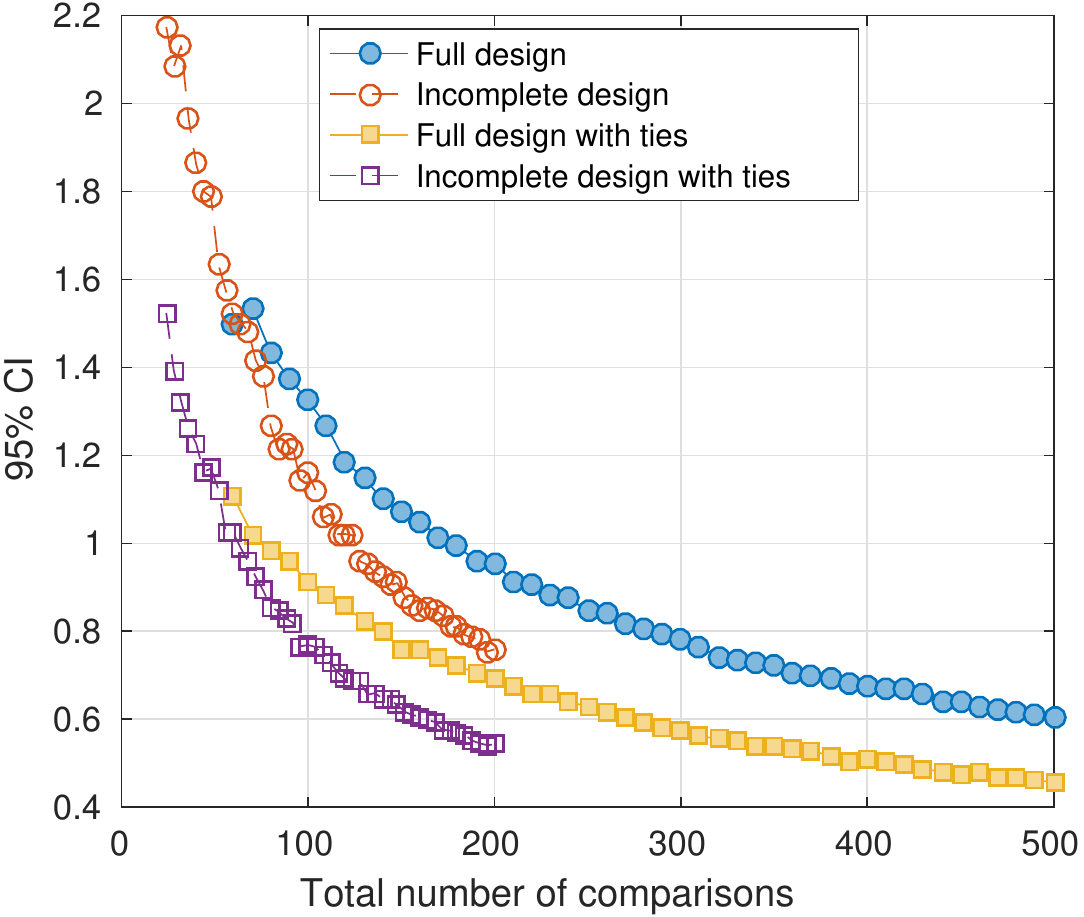}
\includegraphics[width=0.32\textwidth]{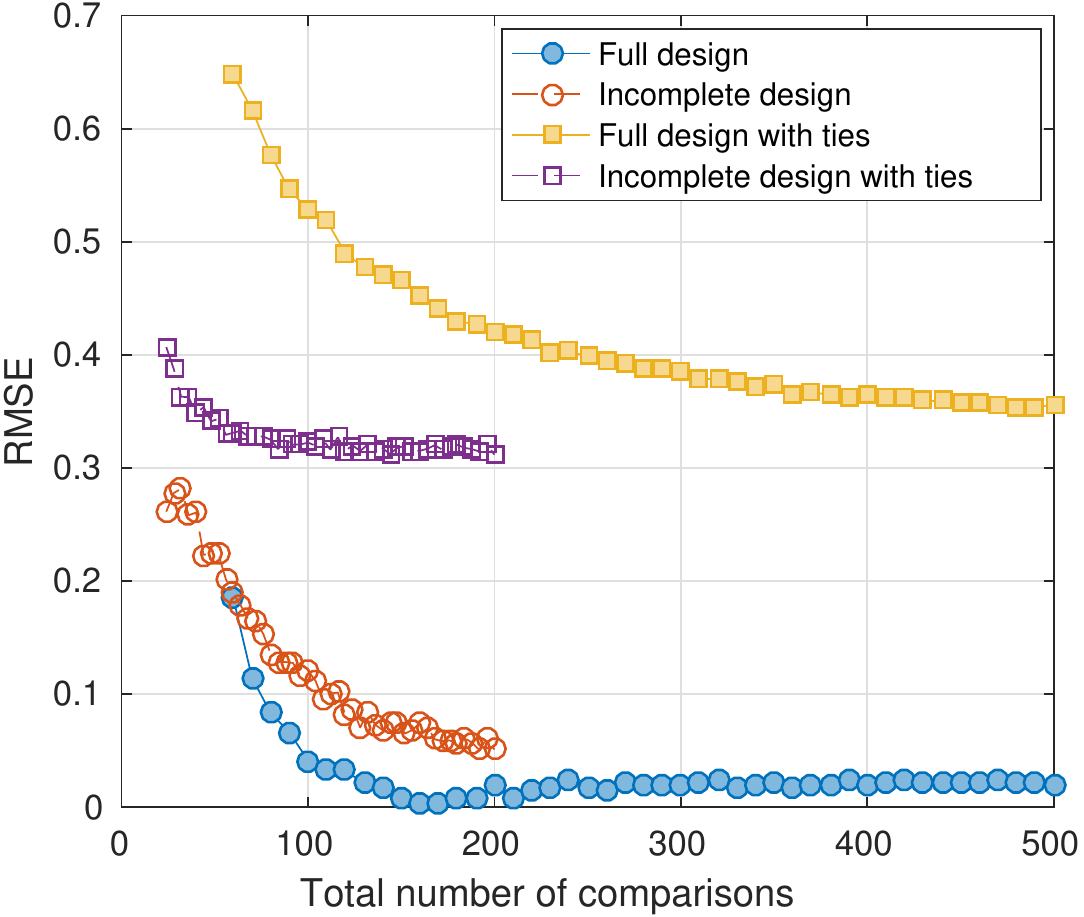}
\caption{The effect of the total number of comparisons on the effect size (left), confidence interval (middle) and RMSE (right). The values computed by simulating 10,000 experiment runs assuming true quality values to be separated by 1 JOD. Note that these results are valid only for those conditions.}
\label{fig:simulated-obs}
\end{figure*}

% \begin{figure*}[t]
% \centering
% \includegraphics[width=0.28\textwidth]{simulation/simulated_obs_d}
% \includegraphics[width=0.28\textwidth]{simulation/simulated_obs_ci}
% \includegraphics[width=0.28\textwidth]{simulation/simulated_obs_rmse}
% \caption{The effect of the total number of comparisons on the effect size (top-left), confidence interval (top-right) and bias (bottom). The values computed by simulating 10,000 experiment runs assuming true quality values to be separated by 1 JOD. Note that this results are valid only for those conditions.}
% \label{fig:simulated-obs}
% \end{figure*}

When designing an experiment we have a choice of comparing all possible pairs of conditions (full or complete design), or only selected pairs, usually those that are the most similar to each other (incomplete design). We are interested in knowing which approach is more efficient and leads to more accurate results. 

We have already shown some results for full and incomplete results in Figure~\ref{fig:simulated-with-prior} when discussing the importance of a prior for small sample sizes. In the plots in that figure incomplete design results in similar accuracy but lower stability in general. However, the plots do not account for the fact that in the full design each participant needs to run many more comparisons. Given $n=5$ compared conditions in our simulation, the full design requires comparing $(5{\cdot}4)/2=10$ pairs, but in the case of incomplete design we compare just 4 pairs: $q_1{\Leftrightarrow}q_2$, $q_2{\Leftrightarrow}q_3$, $q_3{\Leftrightarrow}q_4$ and $q_4{\Leftrightarrow}q_5$.

We replotted the data as the function of the number of comparisons instead of observers in Figure~\ref{fig:simulated-obs}. Please ignore for now ``with ties'' curves and focus on the blue and red lines of full and incomplete designs. The plots show that incomplete design results in more stable and similarly accurate estimates given the same experimental effort. The gain will depend on the number of conditions to compare. For example, if we had 10 conditions, full design would require comparing 45 pairs, but only 9 pairs would need to be compared in incomplete design, resulting in much larger gain. Similar conclusions have been drawn when a sorting algorithm was used \citep{Silverstein2001,Maystre2017}.

% \textcolor{red}{What if each participant has a different variance? We can try to emulate this using different noise for each participant.}
% 
% 
% \textcolor{red}{Can we include a simulation where transitivity does not hold? The main limitation of the simulations performed is all the data is consistent.}
% 
% 
% 
% \textcolor{red}{Can we perform a small experiment showing that with very few observers (e.g. 3, pilot study) we could estimate the order of the comparisons efficiently to decrease later on the number of comparisons?}

%The shortcoming of incomplete design is that we need to know in %advance the likely ordering of quality scores resulting from our %experiment, so that we know which pairs to compare. Alterna

\subsection{Distance between quality scores}

\begin{figure*}[t]
\centering
\includegraphics[width=0.32\textwidth]{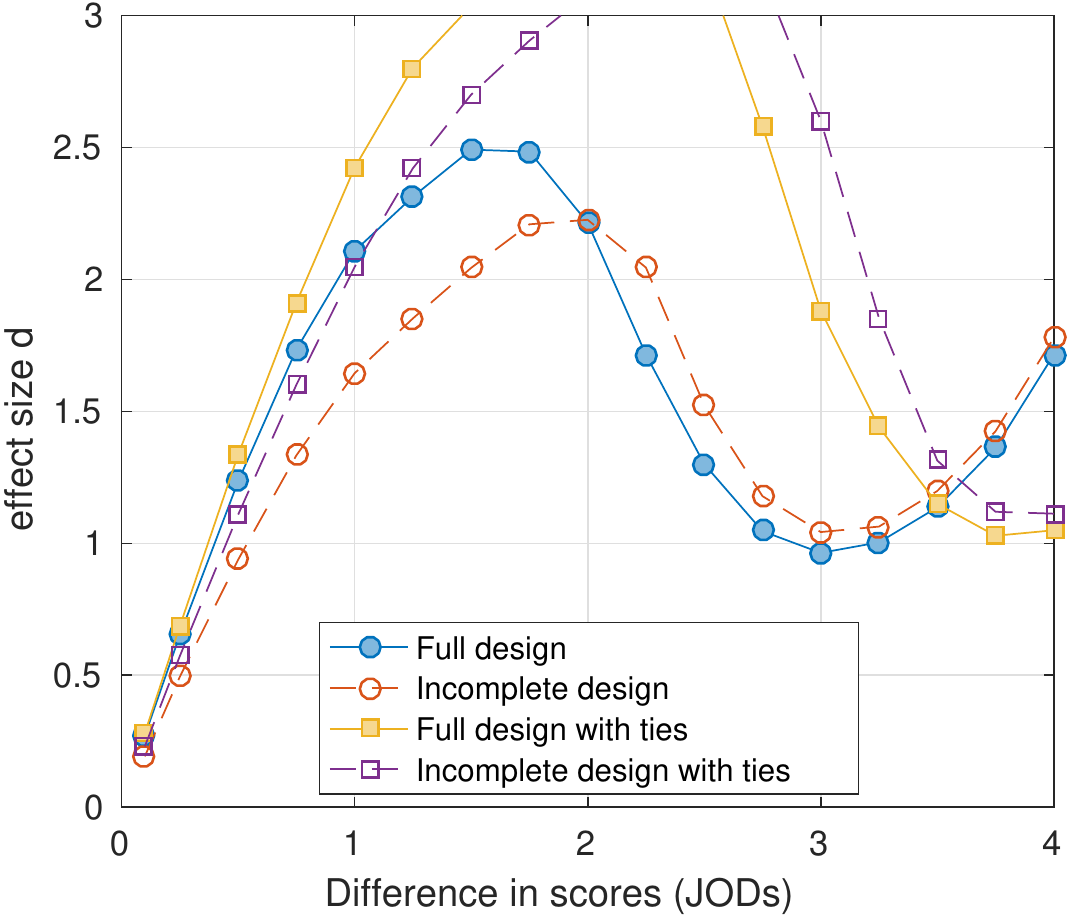}
\includegraphics[width=0.32\textwidth]{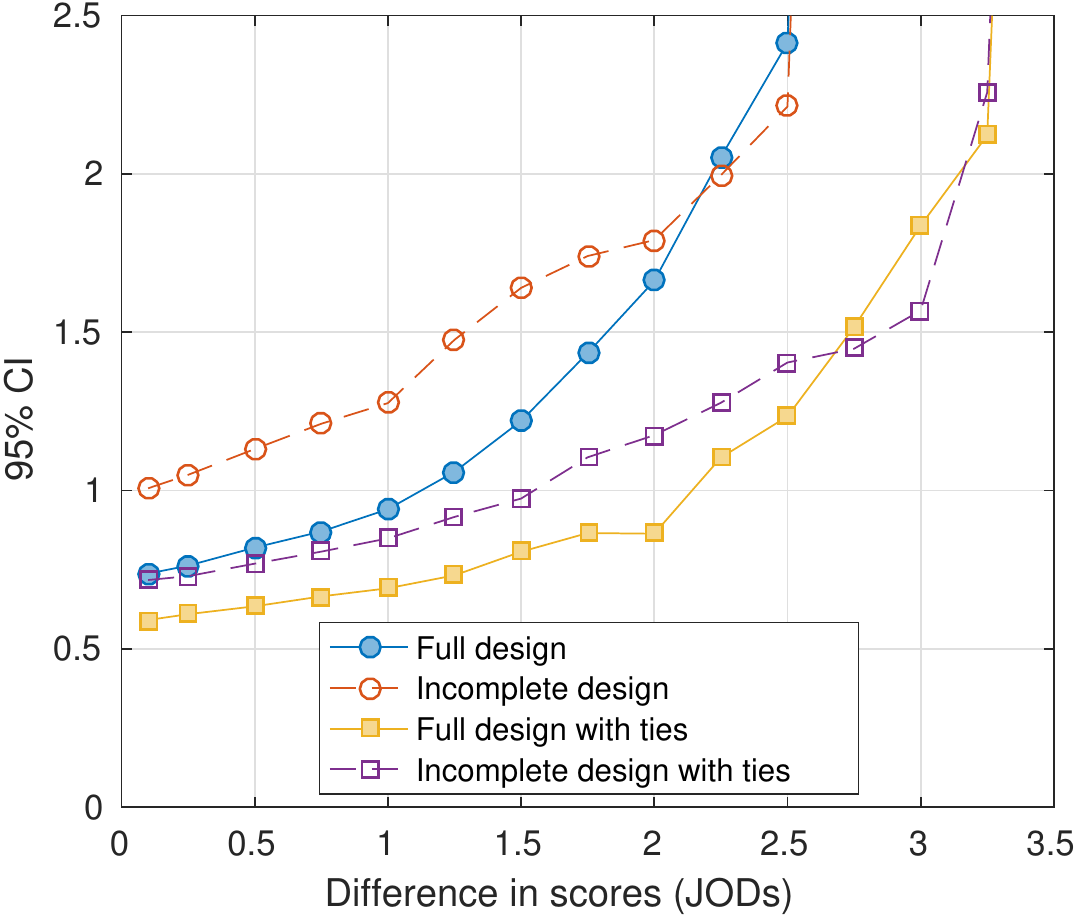}
\includegraphics[width=0.32\textwidth]{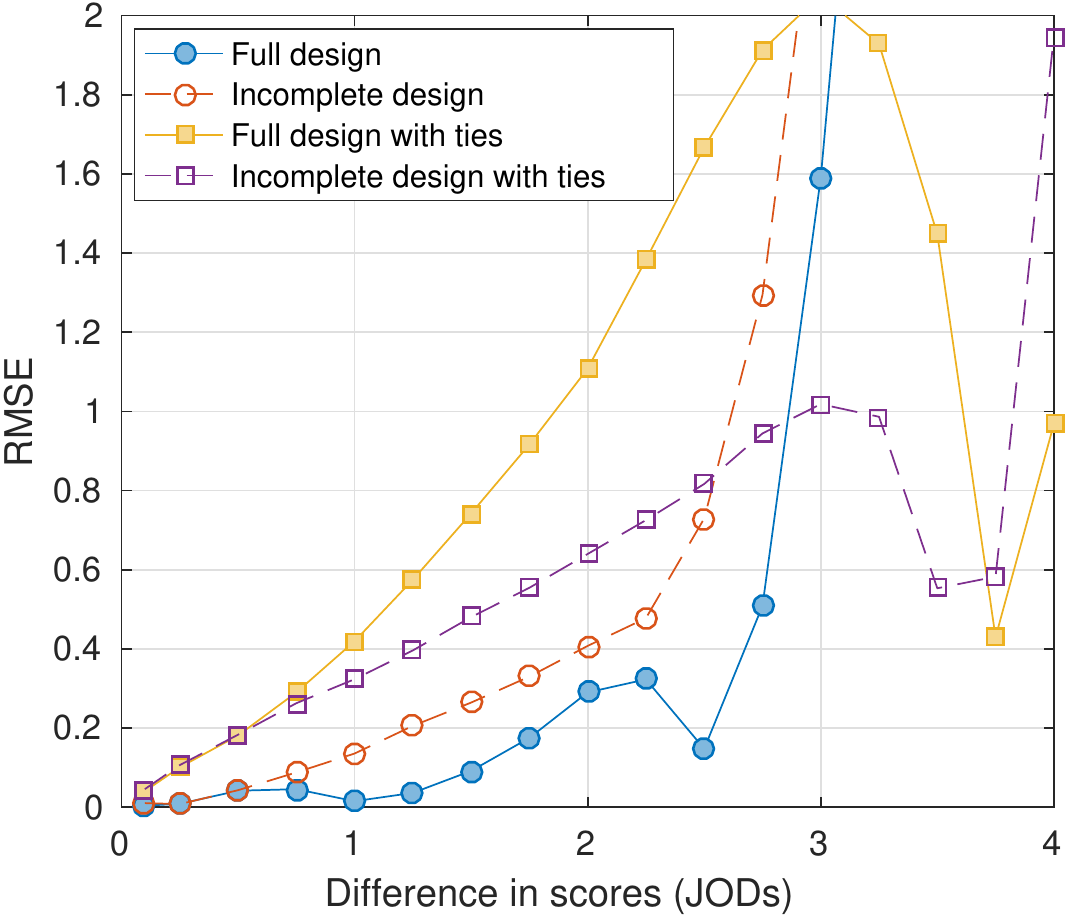}
\caption{The effect of difference between quality scores on the effect size (left), confidence interval (middle) and bias (right). The more measured conditions are separated on the quality scale, the larger is the error and bias. But this also depend whether ties were allowed and whether the all (full design) or only neighbouring conditions (incomplete design) were compared. The values computed by simulating 10,000 experiment runs for 20 observers for each point on the plot. }
\label{fig:simulated-qdiff}
\end{figure*}

The accuracy of scaling methods depends on the distances between quality scores. The scaling becomes especially unreliable if the distance between quality scores is larger than 2 JODs (i.e. $p_{ij}>0.91$). When we suspect that perceptual attributes will be scaled over a larger range than 2 JODS, the difference scaling method \citep{Maloney2003} could be more appropriate. 
To test this effect, we run a Monte-Carlo simulation for different assumed distances (all equal) between true JOD scores, and summarised them in Figure~\ref{fig:simulated-qdiff}. Let us focus on the full design (blue-continuous lines) and ignore all other curves for now. As shown in Figure~\ref{fig:simulated-qdiff}, both the RMSE and confidence intervals increase rapidly, even although the distances between conditions are changed smoothly. It is obvious that measures such as RMSE will increase, since the range of true values increases along x-axis, however, the increase is more abrupt than the linear increase expected.  %confidence intervals get larger as the distance between quality scores increases and peaks at about 2 JODs. But it does not mean that the method is more accurate once the difference are larger than 2 JODs.

%This is because the error (RMSE) starts to increase rapidly. %These figures have been produced taking into account the range of the true quality scale for the metrics of effect size and RMSE.  %Although the range of predicted values (confidence interval) gets smaller above 2 JODs, the absolute predictions on average become worse due to the increased bias. 

\subsection{Experiments with ties}
\label{sec:ties}

Allowing observers to select a third ``no preference'' option when they can not see a difference, is a controversial issue in pairwise comparison experiments, still disputed and researched \citep{Ennis2012,Ennis2012a,Chapman2005,Davidson1970}. 

There are different ways in which ties can be introduced in the statistical analysis \citep{Ennis2012}. For our next experiment we choose the equal-split method: if an observer chooses ``no-preference'', we split the vote in two and add a half-vote to each condition. This may result in a non-integer number of votes, which we round up to the nearest
smaller or larger integer (randomly selected and taking into account the number of comparisons needs to be consistent). We simulate  observers who make the ``no-preference'' choice when the difference between the two conditions is less than a certain threshold. As different observers are unlikely to have the same and consistent opinion when the two conditions are the same, our ``no-preference'' threshold is a random variable $N(0.7,0.3)$ in the space of JOD units. The result of simulating 10,000 experiment runs with ties are compared with  the same experiments but without the tie option in Figures~\ref{fig:simulated-obs} and~\ref{fig:simulated-qdiff}.

Our simulation shows that offering a ``no-preference'' option reduces the size of confidence intervals and improves the effect size. But this happens at the cost of a larger error (see the third plot in Figure~\ref{fig:simulated-obs}). Taking a closer look at the results, we observe that the solution is always under-estimated. There is an intuitive interpretation of this result: offering ``no-preference'' option results in more ``no difference'' responses while the difference is actually there, giving smaller JOD distances and negative bias (under-prediction). The bias is large enough to offset any gains in the reduced confidence intervals. The bias can be potentially eliminated, but it requires modeling the ``no-preference'' selection \citep{Davidson1970} and finding the parameters of that model: how likely will observers select ``no-preference'' where there is actually no difference \citep{Ennis2012a}. This in turn requires collecting extra data: observer responses for two identical conditions. The current version of the \textsf{pwcmp} software does not support modeling ties when scaling, therefore we cannot recommend offering a ``no-preference'' option when this software is used for scaling.

% \subsection{\textcolor{red}{MDS with realworld data}}
% 
% \textcolor{red}{
% How many dimensions do we need to project the data? Can we link this to the number of violations of transitivity?}

\section{Conclusions and limitations}
\label{sec:conclusions}

%This paper presents a guide for using pairwise comparison scaling methods. 
 The choice of pairwise comparison data and scaling methods over a more simplistic analysis presents several advantages: (i) it can be used to compare and rank items that present similar quality (as opposed to direct ordinal rating), (ii) it allows the potential use of incomplete designs to decrease the data to collect (while presenting accurate predictions), (iii) the scaling can be interpreted (especially since the difference measure units can provide information about the probabilities) and (iv) measurement noise in the comparisons can be addressed in a principled way. 
 
 %In this work, we proposed a distance-based prior that improves the estimation and stabilises the results and a new approach to detect outlier observers. Moreover, we provide general guidelines and a toolbox for the common practitioner, addressing topics related to the experimental design. 
 Concerning general guidelines for the experimental design, our results show that incomplete designs can achieve competitive performance if comparisons are appropriately chosen (e.g. neighbours in the quality scale), that the use of ties generally results in an under-estimation of the scaling solution and that it is crucial to ensure that differences between compared conditions are relatively small. Our experiments have also shown the importance of the finite distance prior and screening outlier observers.

The limitation of our work is the assumption of a simple Thurstone Case V observer model, which does not account for dependence between repetitions, observers and conditions, and assumes quality to be explained by a single scalar value.
 
%The limitations of our work are the following: (i) Two different pairwise comparison experiments performed separately can not be compared without doing a third experiment linking both, (ii) the methods proposed here do not account for the fact that different conditions and observers might not be completely independent. %and (iii) our simulated experiments do not account for transitivity violations.

As future work, we would like to extend our analysis and software to include adaptive sampling procedures to reduce the number of required comparisons, and more advanced machine learning techniques that link explanatory variables to the scaling, to facilitate the process of knowledge extraction.
%{Insight on comparisons across different scenes?.} Multi-dimensional scaling and violations of transitivity. 

%   \emph{only} cited references.}
% \medskip
% 
% \small

\bibliographystyle{apalike}

%\bibliography{biblio,IEEEfull,references,mybibfile,biblio2,IEEEfull2}

\end{document}